\documentclass[twocolumn, trackchanges]{aastex7}

\usepackage{longtable}
\usepackage{array}
\usepackage{amsmath}

\newcommand{\lowell}{Lowell Observatory, 1400 W. Mars Hill Rd., Flagstaff, AZ 86001, USA}

\newcommand{\carnegie}{Earth and Planets Laboratory, Carnegie Institution for Science, 5241 Broad Branch Road NW, Washington, DC 20015, USA}

\newcommand{\nau}{Department of Astronomy \& Planetary Science, Northern Arizona University, P.O.\ Box 6010, Flagstaff, AZ 86011, USA}

\newcommand{\planetaryscienceinstitute}{Planetary Science Institute, 1700 East Fort Lowell Rd., Suite 106, Tucson, AZ 85719, USA}

\newcommand{\iaa}{Instituto de Astrof\'isica de Andaluc\'ia, IAA-CSIC, Glorieta de la Astronom\'ia s/n, 18008 Granada, Spain}

\newcommand{\fsi}{Florida Space Institute, University of Central Florida, 12354, Research Parkway, Orlando, FL 32826, USA}

\begin{document}

\title{Rotation and Mutual Eclipse Events Season in the Kuiper Belt: (524366) 2001~XR$_{254}$}

\author[orcid=0000-0002-1506-4248,gname='Audrey', sname='Thirouin']{Audrey Thirouin}
\affiliation{\lowell}
\email[show]{thirouin@lowell.edu}  

\author[orcid=0000-0002-8296-6540,gname='William M.', sname='Grundy']{William M. Grundy}
\affiliation{\nau}
%\affiliation{\lowell}
 \email{w.grundy@gmail.com}
 
\author[orcid=0000-0003-3145-8682,gname=Scott S., sname='Sheppard']{Scott S. Sheppard} 
\affiliation{\carnegie}
\email{ssheppard@carnegiescience.edu}

\author[orcid=0000-0002-6013-9384,gname=Keith S., sname='Noll']{Keith S. Noll} 
\affiliation{\planetaryscienceinstitute}
\email{knoll@psi.edu}

\author[orcid=0009-0006-8584-1416,gname=Jos\'e Mar\'ia , sname='G\'omez-Lim\'on']{Jos\'e Mar\'ia G\'omez-Lim\'on} 
\affiliation{\iaa}
\email{gomezlimon.astro@gmail.com}

 \author[orcid=0000-0002-1788-870X,gname=Ben, sname='Proudfoot']{Benjamin Proudfoot} 
\affiliation{\fsi}
\email{benjamin.proudfoot@ucf.edu}

\begin{abstract}
 
(524366) 2001~XR$_{254}$ is a dynamically Cold Classical Kuiper Belt Object that is a nearly equal-sized wide binary whose upcoming mutual events season makes it a particularly valuable target for physical characterization. In advance of the mutual eclipse events, we conducted a ground-based photometric observing campaign between 2021 and 2026 using the \textit{Lowell Discovery Telescope} to investigate the rotational and physical properties of the system, to refine predictions for its future mutual events, as well as provide a lightcurve we can use as a baseline to identify eclipse events superimposed with the rotation. We derive a double-peaked asymmetric rotational lightcurve with a period of 11.17$\pm$0.04~h and a lightcurve amplitude of 0.42$\pm$0.04~mag. The lightcurve presents a sharp V-shaped minimum consistent with the primary being a close/contact binary, possibly making this a triple system. Using the Keplerian mutual orbit solution, we model the upcoming mutual event season expected between $\sim$2031 and $\sim$2040 and present some individual events for future campaigns. 2001~XR$_{254}$ is one of the few wide binary Kuiper belt systems with a well-determined mutual orbit and a soon observable mutual eclipse event season. Mutual event observations of this possible triple system will provide a rare opportunity to improve component sizes, shapes, densities, and surface properties, offering insights into the formation of the Kuiper Belt.
 
\end{abstract}



%%%%%%%%%%%%%%%%%%%%%%%
%%%%%%%%%%%%% 
\section{Introduction} 
\label{sec:intro}

Kuiper Belt Objects (KBOs), also known as Trans-Neptunian Objects, are relics of the early Solar System that preserve information about the physical conditions and processes that shaped the first planetesimals and the formation of the planets \citep{Morbidelli2020}. Among them, the dynamically Cold Classical population—objects located between Neptune’s 3:2 and 2:1 mean motion resonances on low inclination and low eccentricity orbits—is considered the most primordial subpopulation of the Kuiper belt, as these objects likely formed in situ and experienced little subsequent dynamical excitation \citep{Batygin2011}. One of the defining features of the Cold Classical population is its high fraction of nearly equal-sized, widely separated binary systems \citep{Noll2020}.

Binary and multiple systems are particularly valuable because their mutual orbits provide direct mass measurements. Combined with size estimates, these measurements constrain bulk densities, internal structures, and compositions, offering key diagnostics for distinguishing between formation pathways such as gravitational collapse and collisional evolution \citep{Noll2020, Grundy2019}. To date, only about 39 of the $\sim$ 120 resolved binary systems known in the Kuiper belt have fully determined mutual orbits and system masses \citep{Grundy2019}. However, the individual sizes and shapes of the components still remain poorly constrained for most systems.

Stellar occultations can provide precise measurements of object sizes, shapes, and albedos, and can also reveal rings and/or satellites \citep{Ortiz2020}. However, such events require dense multi-chord coverage along a narrow visibility path, and, in the case of wide binaries, simultaneous occultations by both components are exceedingly unlikely (but simultaneous detections of the components are not required as several occultations of the primary and secondary allow for the characterization of both) \citep{Sickafoose2019}. Mutual events—eclipses and occultations between binary components—offer one of the most powerful methods for characterizing binary and multiple systems. These events can constrain component sizes, shapes, densities, albedos, orbital parameters, and even surface color variations maps \citep{Binzel1997, Descamps2007, Scheirich2009, Benecchi2014, Rabinowitz2020, Berthier2020}. Unlike stellar occultations, mutual events are observable from any location on Earth and allow both components of the system to be studied simultaneously throughout an event. If the mutual orbits of a binary are well known, mutual event seasons, which happen when the orbits of the binary components are viewed nearly edge-on, can be predicted years ahead of time. These eclipse and occultation events between the mutual components of a binary system can be observed multiple times over a few years, allowing repeated observations to obtain a detailed analysis of the components \citep{Proudfoot2026, Thirouin2025}. However, mutual event seasons are rare, and only a handful have been observed in the Kuiper belt to date \citep{Binzel1997, Benecchi2014, Rabinowitz2020}.

Several binary and multiple systems in the Kuiper belt are currently undergoing, or are predicted to soon undergo, mutual events seasons \citep{Proudfoot2026, Thirouin2025, Grundy2019}. In this work, we focus on the known resolved wide binary system 2001~XR$_{254}$, in the Cold Classical subpopulation, whose next mutual event season is expected to begin within the next decade. In Section~\ref{sec:presentation}, and Section~\ref{sec:orbit} we summarize the known properties and Keplerian mutual orbit parameters of the system. In Section~\ref{sec:observations}, we present its rotational and physical characteristics derived from ground-based observations. In Section~\ref{sec:model}, we model the system assuming the derived mutual orbit and predict its upcoming mutual event season (Section~\ref{sec:season}).

%%%%%%%%%%%%%%%%%%%%%%%
%%%%%%%%%%%%% 
\section{Current Knowledge of \\ (524366) 2001~XR$_{254}$} 
\label{sec:presentation}

The Kuiper Belt Object (524366) 2001~XR$_{254}$ was discovered in December 2001 using the 2.2~m University of Hawaii telescope \citep{Sheppard2002discovery}. With a semi-major axis$\footnote{Orbital elements were retrieved from the Minor Planet Center (MPC) in May 2026: \url{https://minorplanetcenter.net/db_search/show_object?object_id=524366}}$ of 43.186~au, an eccentricity of 0.029, and an inclination of 1.24$^{\circ}$, 2001~XR$_{254}$ is classified as a dynamically Cold Classical object within the main Kuiper belt located between Neptune’s 3:2 and 2:1 mean motion resonances.

Observations obtained with the \textit{Hubble Space Telescope (HST)} in December 2006 revealed a distant satellite orbiting 2001~XR$_{254}$ \citep{Noll2008}. The magnitude difference between the two components, $\Delta{mag}=0.415\pm0.064$~mag, indicates that the system is a nearly equal-sized binary \citep{Grundy2009}. 

Using \textit{HST} observations in the F606W and F814W filters, \citet{Benecchi2009} measured nearly identical surface colors for the two components: F606W$-$F814W$=0.74\pm0.09$~mag for the primary and $0.76\pm0.10$~mag for the secondary, corresponding to a color difference of only $-0.02\pm0.13$~mag. These measurements indicate that both components have similar surface compositions$\footnote{\citet{Benecchi2009} converted the \textit{HST} F606W$-$F814W colors into the Johnson-Cousins system, obtaining $(V-I)_{primary}=1.04\pm0.11$~mag and $(V-I)_{secondary}=1.06\pm0.12$~mag.}$ \citep{Benecchi2009}. Although dynamically Cold Classical KBOs are typically ultra-red, the components of 2001~XR$_{254}$ exhibit only moderately red surfaces, leading \citet{Fraser2017} to classify the system as a blue binary. However, due to the large uncertainty on its visible spectral slope, the system was a borderline blue/red binary. Recent \textit{James Webb Space Telescope} spectroscopic observations revealed that 2001~XR$_{254}$ is a typical Cold Classical object based on its spectrum \citep{Wong2025}.  

From \textit{HST} observations obtained between 2006 and 2007, \citet{Grundy2009} derived the first mutual orbit solution for the 2001~XR$_{254}$ wide binary system. This solution was later refined by \citet{Grundy2019} using additional \textit{HST} and \textit{Keck~II} observations. This has been most recently updated by \citet{Proudfoot2026} through a combined analysis of all available \textit{HST} and \textit{Keck~II} astrometry spanning 2006--2024. \citet{Proudfoot2026} reported tentative evidence for orbital precession, suggesting that the mutual orbit may be non-Keplerian. However, the Keplerian solution could only be rejected at the 2.4$\sigma$ level, and additional observations are required to confirm or rule out orbital precession. 

Based on \textit{Herschel Space Observatory} data, \citet{Vilenius2012} determined the radiometric sizes and albedos of several binary systems in the Kuiper belt, including 2001~XR$_{254}$. Combined with the system's mass derived by \citet{Grundy2009}, they found that the system's radiometric diameter is 200$^{+49}_{-63}$~km for a geometric albedo of 0.17$^{+0.19}_{-0.05}$ and a bulk density of 1.4$^{+1.3}_{-1.0}$~g/cm$^{3}$. The diameter of the primary is estimated at D$_{primary}$=170$\pm$43~km, and the satellite's diameter to D$_{satellite}$=141$\pm$36~km, assuming the albedo derived using the \textit{Herschel Space Observatory} data as well as spherical shape for the two components \citep{Proudfoot2026, Vilenius2014, Vilenius2012}. 

\section{Keplerian mutual orbit determination}
\label{sec:orbit}

Since non-Keplerian orbital motion was not convincingly detected by
\citet{Proudfoot2026}, we used a simpler Keplerian mutual orbit.  This
orbit solution was fitted to the same \textit{HST} and \textit{Keck~II}
data set spanning years from 2006 to 2024 that was already described in
\citet{Proudfoot2026}. We employed the same methods as in previous publications
\citep[e.g.,][]{Grundy2019} to iteratively adjust orbital parameters
to minimize the chi-squared statistic between observed and calculated
relative positions of the two bodies, accounting for the shifting
parallaxes caused by the motion of Earth and 2001~XR$_{254}$ around
the Sun. Uncertainties in the orbital elements were estimated via the
Monte Carlo bootstrap method. The resulting elements and uncertainties
are listed in Table~\ref{tab:orbitalelements}. In this Table we also display the orbital elements from \citet{Proudfoot2026} MCMC fits. The ($a$, $e$, $i$, $\Omega$, $\omega$, $\mathcal{M}$) values in Table~A2 of \citet{Proudfoot2026} are the osculating elements of their non-Keplerian solution at the stated reference epoch. Since osculating elements are well-defined at any instant independent of the perturbing force model, we converted these directly into our parametrization and reference frame for the comparison in Table~\ref{tab:orbitalelements}. We do not show a period for this orbit, since it would not be directly comparable to the period of the Keplerian solution, which blends the orbital period with precession. Likewise, we do not show the mass for the Keplerian solution, as it would be computed using that blended period.

\begin{table*}[]
\centering
 \caption{Mutual orbit parameters based on \textit{HST} and \textit{Keck~II} data from 2006 to 2024. \cite{Proudfoot2026} fit is also shown for comparison. Angles are referenced to the International Celestial Reference Frame (ICRF).}
 \label{tab:orbitalelements}
\begin{tabular}{|l|l|l|}
\hline
Orbital elements      & This work  & \cite{Proudfoot2026}       \\ \hline
Orbital period [days] & 125.647$\pm$0.003 & -  \\ \hline
System mass [kg] & -  & (4.037$\pm$0.054)$\times 10^{18}$ \\ \hline
Semi-major axis [km] & 9244$\pm$21 & 9299$^{+43}_{-39}$ \\ \hline
Eccentricity & 0.544$\pm$0.003 & 0.552$^{+0.005}_{-0.004}$ \\ \hline
Inclination [$^\circ$]  & 41.55$\pm$0.14 & 41.15$^{+0.19}_{-0.19}$ \\ \hline
Mean longitude [$^\circ$]  & 151.30$\pm$0.20 & 151.59$^{+0.25}_{-0.26}$ \\ \hline
Longitude of ascending node [$^\circ$]  & 340.64$\pm$0.23 & 340.96$^{+0.29}_{-0.26}$ \\ \hline
Longitude of periapsis [$^\circ$]  & 244.7$\pm$0.5 & 245.83$^{+0.76}_{-0.78}$ \\ \hline
Epoch [Julian date]& 2454300.0 & 2454300.0  \\ \hline
\end{tabular}
\end{table*}

The Keplerian and non-Keplerian mutual orbits, from this work and from \citet{Proudfoot2026}, produce orbital elements compatible within $~1.5\sigma$ for the epoch JD 2454300.0.
With the non-Keplerian mutual orbit, \citet{Proudfoot2026} showed that 2001~XR$_{254}$ is expected to undergo mutual events starting in $\sim$2031 to the 2040s, which is in agreement with the season's midpoint estimated by \citet{Grundy2019} with an earlier version of the Keplerian mutual orbit. In this work, we will update \citet{Grundy2019} prediction for the upcoming mutual events season, obtaining predictions also compatible with \citet{Proudfoot2026}.

%%%%%%%%%%%%%%%%%%%%%%%
%%%%%%%%%%%%%

\section{Rotational Study: Observations and Analysis} 
\label{sec:observations}

\subsection{Observing runs}

We conducted an observing campaign with the \textit{Lowell Discovery Telescope (LDT)} between 2021 and 2026 to obtain the first ground-based rotational lightcurve of 2001~XR$_{254}$. 

The 4.3~m \textit{LDT}, located near Happy Jack in Arizona, was used with the Large Monolithic Imager (LMI), which provides a 12.3$\arcmin \times$12.3$\arcmin$ field of view at a pixel scale of 0.12$\arcsec$/pixel (unbinned). To maximize the signal-to-noise ratio while minimizing fringing, observations were acquired with a broad-band VR filter. Individual exposure times ranged from 450 to 600~s, depending on weather and seeing conditions. Every observing night, we obtained bias and dome flat frames to create median bias and median dome flat frames in the VR-filter for calibration. In total, we present in this work a set of ten observing nights carried out over five years (Table~\ref{tab:Obs_log}).  
 
%\startlongtable
\begin{deluxetable}{cccc}
% \tabletypesize{\scriptsize}
\tablecaption{\label{tab:Obs_log} 2001~XR$_{254}$ observing log for our 2021-2026 campaign. Heliocentric and geocentric distances as well as phase angles are reported for each observing night. Ephemeris were computed by the Minor Planet Center (MPC). We used the VR-filter at the \textit{Lowell Discovery Telescope (LDT)} for all observing nights.  \\  }
\tablewidth{0pt}
\tablehead{  UTC  &  $\Delta$   &    r & $\alpha$      \\
       Observing night       &  [AU]  &  [AU]  &  [$^{\circ}$]      }
\startdata
2021.03.05      &43.126 & 43.876   &0.9  \\
2021.12.03        &43.286 & 43.859  &1.1   \\
2021.12.05     & 43.258 & 43.859  &1.0  \\
2022.03.26       &43.358  &43.852 &1.1   \\
2022.12.19      &43.202 & 43.836 & 1.0   \\
2023.12.13      &43.147 & 43.815 &1.0 \\
2024.02.13       &42.863&  43.811&0.4 \\  
2025.11.25      &43.394 & 43.770 &1.2 \\
2026.02.15        &42.816 & 43.765&0.4\\
2026.03.25        &43.182 & 43.763  & 1.1  \\
 \hline
\hline
\enddata
\end{deluxetable}

\subsection{Data reduction, analysis, and lightcurve interpretation}

Below, we summarize the main steps of our data reduction and analysis; this is similar to our past observations and reductions and a more detailed description can be found in \citet{Thirouin2013}. All science images were calibrated using median bias and dome-flat frames. After calibration, the flux of 2001~XR$_{254}$ and comparison stars was measured using aperture photometry \citep{Stetson1987}. The optimal aperture size was determined using the growth-curve technique described in \citet{Howell1989}, which provides the best compromise among maximizing the signal-to-noise ratio, minimizing the sky background, and retaining most of the target flux. Images affected by cosmic-ray hits or contamination from nearby stars were discarded. Particular care was taken to avoid spurious photometric measurements caused by background contamination within the aperture.
 
Because our observations span multiple years, all photometric measurements were corrected for light-travel time before analysis. Periodic signals in the light-time corrected time-series photometry were then searched for using the Lomb periodogram technique \citep{Lomb1976}. The strongest (i.e., tallest) peak in the Lomb periodogram provides the most likely periodicity, but the corresponding lightcurve must then be evaluated to determine whether it is best described as a single-peaked or double-peaked lightcurve. In the case of 2001~XR$_{254}$, the phased lightcurve is asymmetric, with one minimum significantly deeper than the other. Therefore, the rotational period of 2001~XR$_{254}$ is twice the periodicity identified by the Lomb periodogram. We also point out that the Lomb periodogram may have several peaks above the 99.9\% confidence level. In such a case, additional data can be required to discard aliases, but an inspection of the lightcurves produced by these aliases' period can also be performed to discard their likelihood. 

The morphology and amplitude of a rotational lightcurve provide constraints on an object’s rotational and physical properties. A low amplitude lightcurve ($\Delta m \lessapprox$0.2~mag) is generally consistent with a spheroid or nearly spheroidal object (also known as MacLaurin spheroid) or with an object viewed (close to) pole-on. A sinusoidal lightcurve with an amplitude between $\sim$0.2 and $\sim$0.4~mag is typically associated with a rotating elongated triaxial body (Jacobi ellipsoid) not observed pole-on. In the case of an elongated object, and thus a sinusoidal lightcurve, a Fourier series fit can be used to fit the lightcurve. In the case of a double-peaked lightcurve, a second-order Fourier series can be used to fit the observed lightcurve. A second-order Fourier fit corresponds to the equation: \begin{multline}
Fit_{Fourier} = a + b  \cos(2\pi \phi_{rot}) + c \sin(2\pi \phi_{rot}) \\
+ d  \cos(4 \pi \phi_{rot}) + e \sin(4 \pi \phi_{rot})
\end{multline}
where a, b, c, d, and e are constants and $\phi_{rot}$ is the rotational phase. In contrast, a non-sinusoidal lightcurve characterized by broad, inverted U-shaped maxima and sharp V-shaped minima may indicate shadowing effects typical of an object that is a contact/close binary. \citet{Leone1984, Chandrasekhar1987} showed that a nearly equal-sized contact binary (Roche binary) observed equator-on typically produces a lightcurve amplitude larger than 0.9~mag, along with characteristic V-shaped minima and U-shaped maxima. However, when a close/contact binary is observed at lower aspect angles (equator-off), the lightcurve amplitude decreases and these distinctive features from shadowing effects of the two components can become less pronounced or even smoothed out \citep{Lacerda2011}. As a result, the 0.9~mag threshold is not always reached, since the observed lightcurve depends strongly on the viewing geometry, which changes over the orbit of an object. A more detailed lightcurve interpretation and analysis is available at \citet{Sheppard2002, Lacerda2007, Sheppard2008, Lacerda2011, ThirouinSheppard2022, ThirouinSheppard2024}, among others.

\subsection{Lightcurve of 2001~XR$_{254}$}
\subsubsection{Lightcurve: presentation and interpretation}

In this section, we present the ground-based rotational lightcurve of 2001~XR$_{254}$ and its interpretation. We note that, as we are dealing with a binary system, several periods: period of the primary, period of the satellite, and the orbital period, all have to be taken into account. The orbital period is known to be about 125~days (Table~\ref{tab:orbitalelements}, Section~\ref{sec:orbit}), but the rotational periods of the system's components are unknown. The \textit{HST} data are too sparse to retrieve the separated full lightcurves of each component, but they give us constraints on the components' variabilities. Using all available \textit{HST} images, we estimated that the satellite shows a variability of about 0.1-0.3~mag while the primary shows a bit more variability at $\sim$0.4-0.5~mag.

\begin{figure}
  \includegraphics[width=9cm,angle=0]{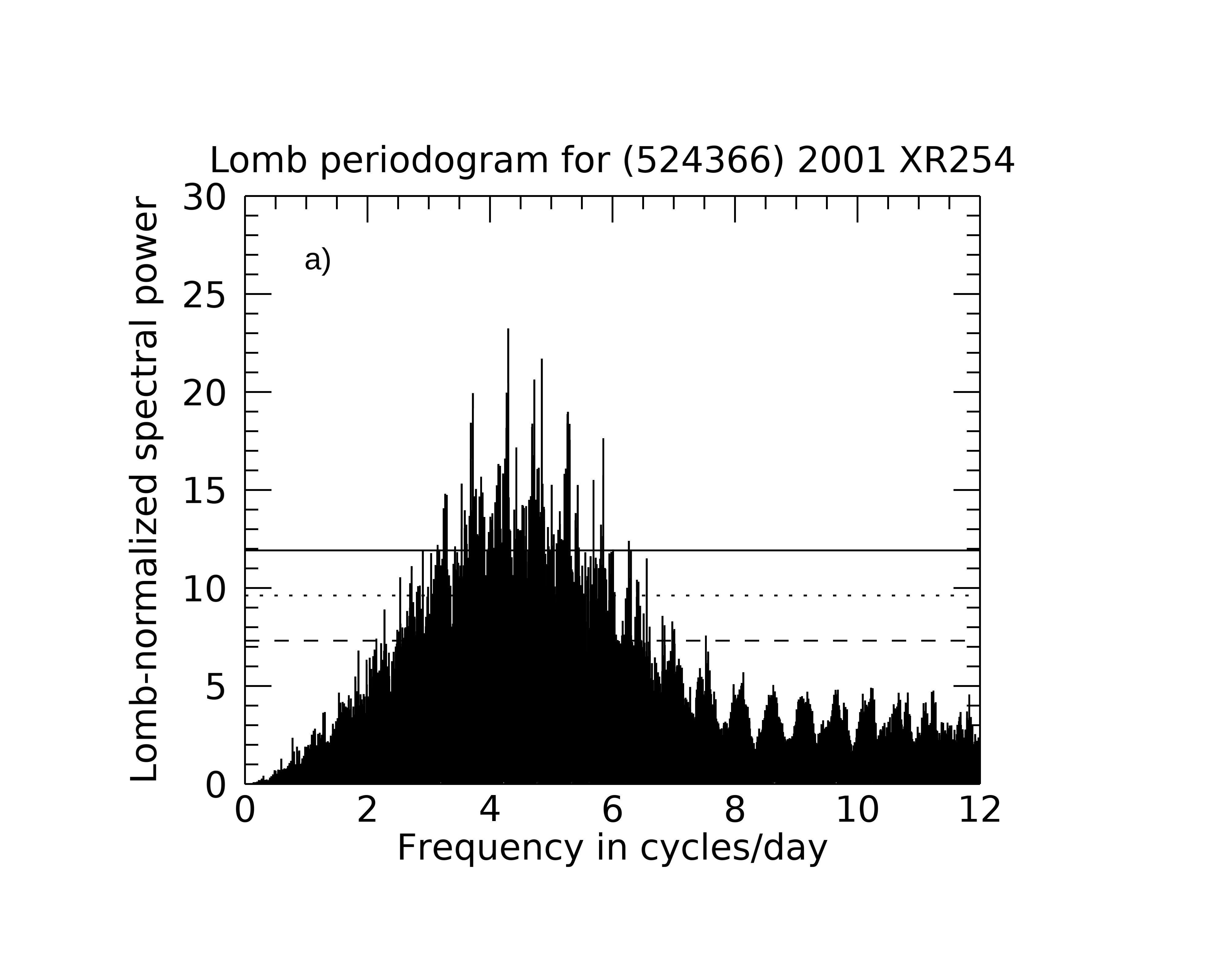} 
 \includegraphics[width=9cm,angle=0]{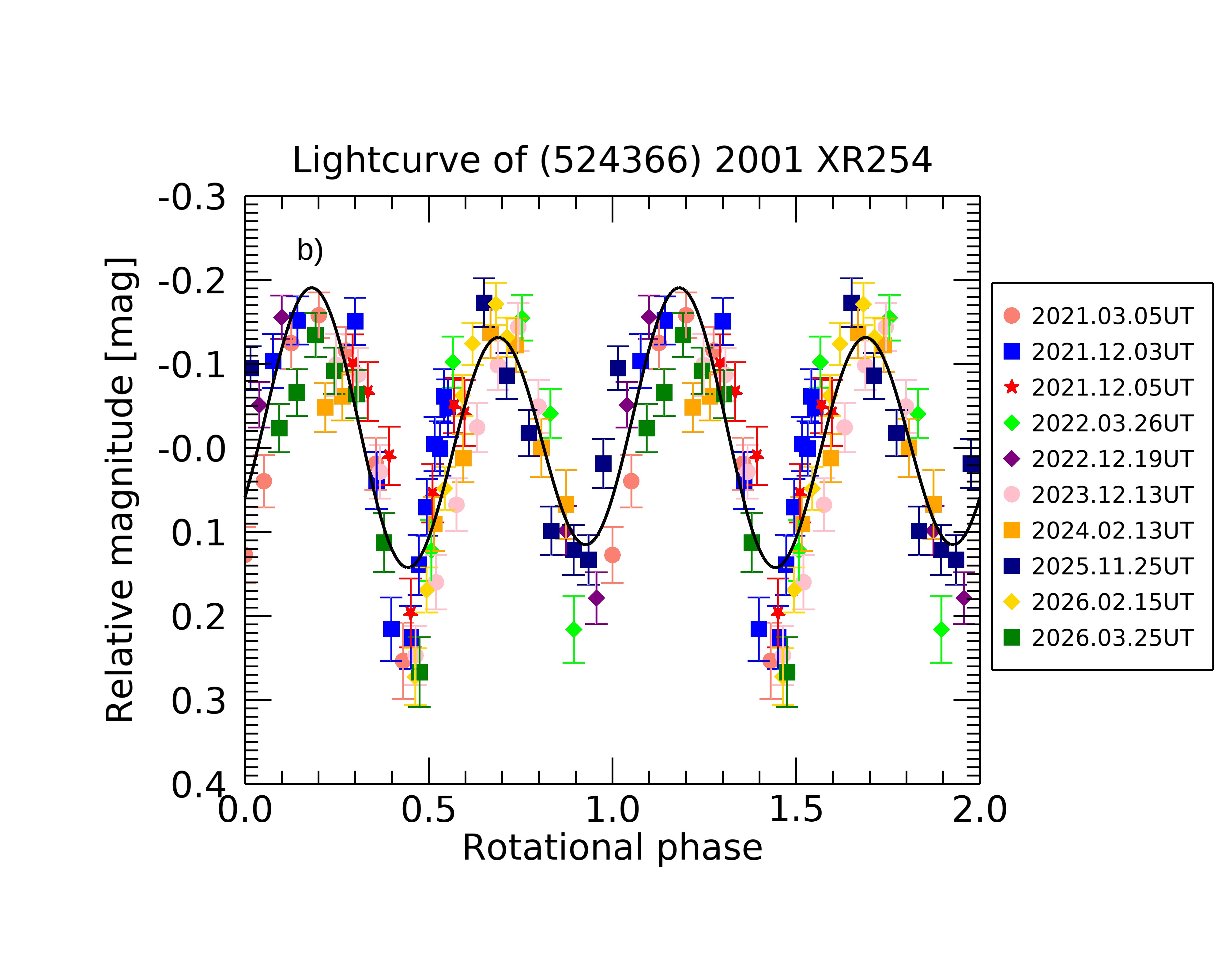}  
\caption{The main peak of the Lomb periodogram (plot a)) is located at 4.298~cycles/day. Horizontal lines overplotted in plot a) correspond to confidence levels of 90\% (dashed line), 99\% (dotted line), and 99.9\% (continuous line). The double-peaked lightcurve of 2001~XR$_{254}$ with a rotational period of 11.17$\pm$0.04~h has an amplitude of 0.42$\pm$0.04~mag (plot b)). The black continuous line in plot b) is a second order Fourier Series fit, that does not match the minimum very well. We attribute most of the observed short-term variability to the primary component of the system. }
\label{fig:XR254}
\end{figure}

First, we examined the data from each year separately and detected no amplitude variations resulting from changes in the system's geometry. Consequently, we treated the entire dataset as a single, unified dataset. A search for periodic signals in the light-time corrected ground-based photometry shows that the strongest peak in the Lomb periodogram occurs at 4.298~cycles/day, corresponding to a single-peaked period of 5.58~h (Figure~\ref{fig:XR254}). However, the phased lightcurve is asymmetric, with the first minimum approximately 0.1~mag deeper than the second. A single-peaked solution cannot reproduce this asymmetry and instead favors a double-peaked lightcurve. We therefore adopt a true rotational period of 11.17~h, twice the periodicity identified by the Lomb periodogram. The corresponding double-peaked lightcurve is shown in Figure~\ref{fig:XR254}. Such an asymmetry can be due to albedo spot(s) on the object surface or an irregular shape. We attribute the lightcurve in Figure~\ref{fig:XR254} to the primary of the system. Therefore, we infer that the rotation of the primary is 11.17~h and its variability is 0.42$\pm$0.04~mag.

Since the lightcurve amplitude is higher than 0.2~mag, this suggests the object is not perfectly spheroidal and/or observed pole-on. We attempted to fit the lightcurve with a Fourier series model, as shown in Figure~\ref{fig:XR254}. However, the fit ($\chi^2$ = 5.31) fails to reproduce the sharp V-shaped minimum and broad maximum observed in the data. This mismatch suggests that the lightcurve is not consistent with the smooth, sinusoidal variation expected from a rotating triaxial ellipsoid. As mentioned, the first minimum shows a V-shape, and the maximum is broad, making 2001~XR$_{254}$ a candidate for the primary being a close/contact binary based on the lightcurve morphology from shadowing effects of the two close components. As the 0.9~mag threshold is not reached, it may be that the object is not currently being viewed equator-on or that the contact binary size ratio of the two components is less than one as both of these situations would result in a lower amplitude lightcurve \citep{Lacerda2011}. Future observations a few years from now at different viewing geometries are needed to better understand the nature of the primary.

Finally, we find no strong evidence for an additional periodic signal that could be associated with the satellite, preventing us from constraining its rotational period. Because the rotational period of the primary differs from the orbital period, we conclude that the system is asynchronous. 

\subsubsection{Lightcurve: modeling with \texttt{Candela}}

To confirm our lightcurve interpretation, we used the \texttt{Candela} software described in \citet{Thirouin2025}. To summarize, \texttt{Candela} is a forward-modeling software developed to simulate mutual events in binary and multiple small body systems, as well as lightcurves of individual objects. The code uses the mutual orbit parameters together with the rotational period, size, shape, and photometric properties of each component to compute synthetic lightcurves and predict the timing, duration, and depth of mutual events. By accounting for changing viewing geometry over time, \texttt{Candela} can model entire mutual event seasons and explore how uncertainties in physical or rotational parameters affect the predicted lightcurves. We can also provide an observed lightcurve to the \texttt{Candela} software to retrieve the object that best fits the lightcurve. 

\begin{figure}
  \includegraphics[width=9cm,angle=0]{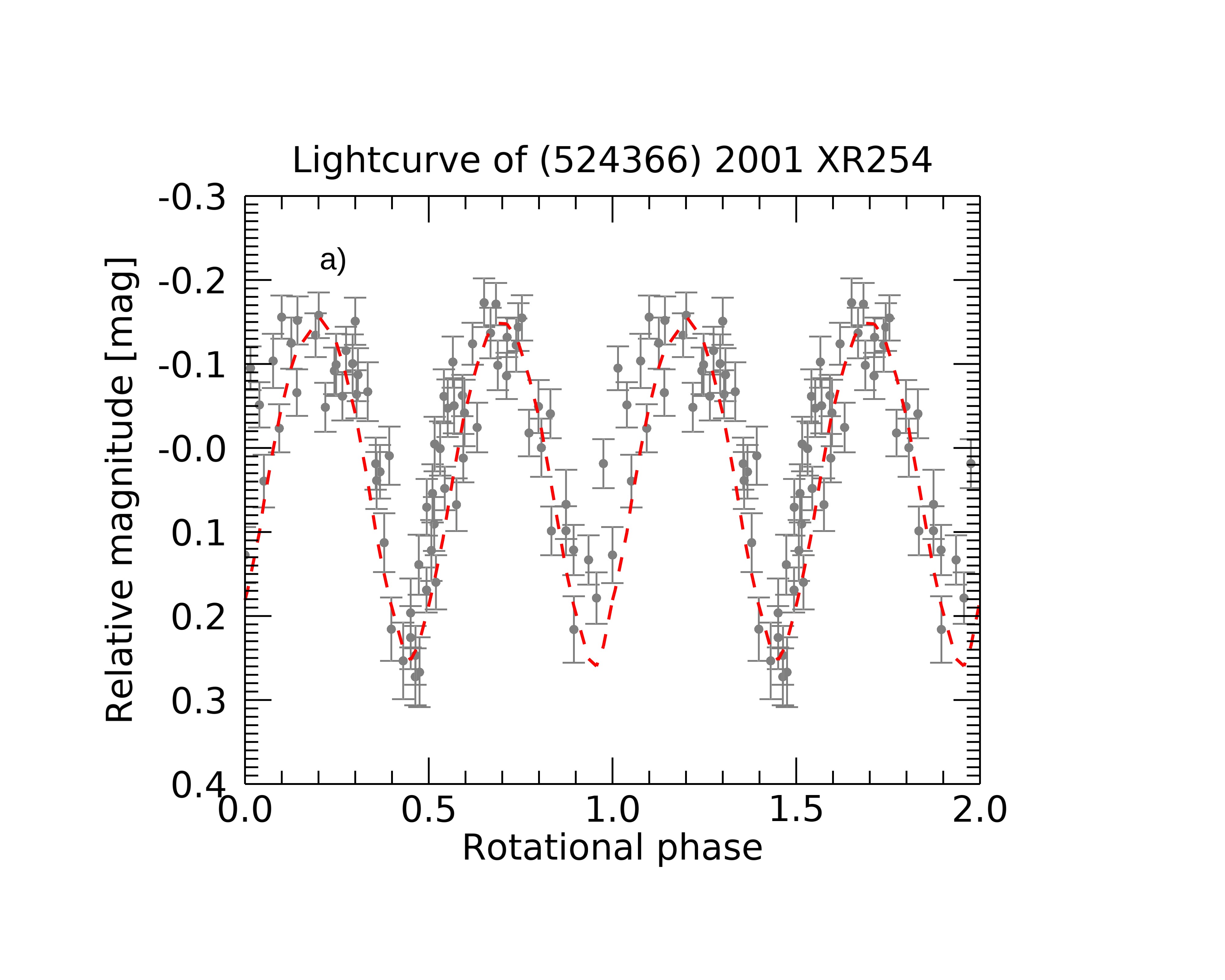} 
\includegraphics[width=9cm,angle=0]{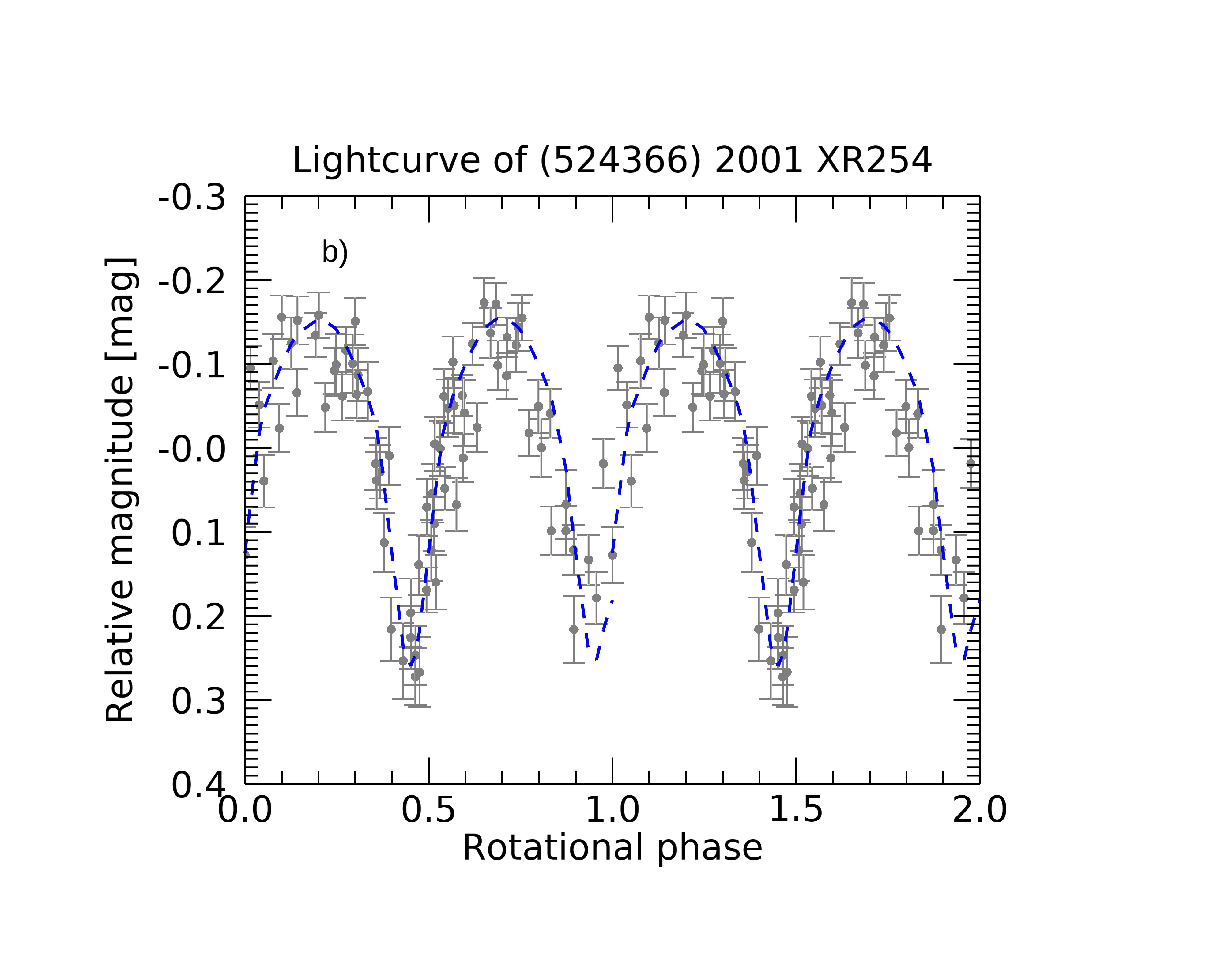} 
\caption{The overplotted red discontinuous line in plot a) corresponds to the best predicted lightcurve assuming a triaxial ellipsoid with semi-major axes such as a=94~km, b=93~km, and c=44~km, a density of 1.4~g~cm$^{-3}$, and an albedo of 0.17. In plot b), the blue discontinuous line is the predicted lightcurve for a contact binary with a mass ratio, m$_{B}$/m$_{A}$ of 0.2, with the component~A of 71$\times$67$\times$59~km while the component~B is 47$\times$36$\times$33~km. All observational data (from Figure~\ref{fig:XR254}) are now plotted as gray dots for clarity. }
\label{fig:XR254Candela}
\end{figure}

As a first step, we use \texttt{Candela} to model the primary component and determine the physical parameters that best reproduce the observed lightcurve shown in Figure~\ref{fig:XR254}. We will test the triaxial ellipsoid and the close/contact binary options. We note that we are considering a perfect ellipsoid and contact binary shapes for this work, and we are not considering irregular shapes with concavities/craters on the surface. To reduce the parameter space, we adopt fixed values for the density and albedo of 1.4~g~cm$^{-3}$ and 0.17 (values derived by \citet{Vilenius2012}), respectively, and assume an equator-on viewing geometry. Using the observed lightcurve, these fixed parameters, and a range of trial values for the semi-axes of a triaxial ellipsoid ($a>b>c$), assuming rotation about the shortest axis ($c$), \texttt{Candela} searches for the combination of shape parameters that best reproduces the observed photometric variability. Under this assumption, the best-fit solution corresponds to semi-axes of $a=94$~km, $b=93$~km, and $c=44$~km if we assume that the lightcurve is produced by an ellipsoid. In Figure~\ref{fig:XR254Candela} (Plot a)), we plotted the best predicted lightcurve (red discontinuous line) and the photometric data corresponding to the observed lightcurve we previously presented. The predicted lightcurve is not a perfect match to the observed lightcurve with a $\chi^{2}$=1.88. In fact, the broad maxima and sharp minima of the observed lightcurve are not well reproduced by the predicted one. 

As a second step, we use \texttt{Candela} to retrieve the best parameters for a close/contact binary configuration to explain the observed lightcurve. We used the same fixed parameters as for the ellipsoid case. \texttt{Candela} favored a close/contact binary with a mass ratio of 0.2 between the component~A and the component~B (where mass ratio = m$_{B}$/m$_{A}$). The shape of the component~A is 142$\times$134$\times$118~km (full axis) while the component~B is 94$\times$72$\times$66~km (full axis); axis-size uncertainties are estimated to be $\pm$5$\%$. However, we emphasize that we have not taken into account the error bars on the density, albedo, and objects' sizes, which may increase the uncertainty. The predicted contact binary lightcurve gives a $\chi^{2}$=1.09, which is a better fit than the triaxial ellipsoid case (Plot b) in Figure~\ref{fig:XR254Candela}). We note that the second minimum of the observed lightcurve is deeper than the first, which may be due to small albedo spot(s). \texttt{Candela} currently does not include albedo spots in the model and thus this option will be part of future work. For the rest of this work, we will consider that the primary is a close/contact binary with a mass ratio of 0.2. 

%%%%%%%%%%%%%%%%%%%%%%%
%%%%%%%%%%%%%
  
\section{Modeling of the system} 
\label{sec:model}

As shown in the previous section, the primary is best modeled as a close/contact binary composed of a component A with dimensions of 142$\times$134$\times$118~km and component B with dimensions of 94$\times$72$\times$66~km. We emphasize that this represents our preferred model under the assumptions of an albedo of 0.17, a density of 1.4~g~cm$^{-3}$, and an equator-on viewing geometry since the mutual orbit is nearly edge-on. This analysis does not account for the uncertainties in the density and albedo reported by \citet{Vilenius2012}, nor does it explore alternative viewing geometries. As a result, the solution presented here should be regarded as one plausible best-fit model rather than a unique physical interpretation.

In contrast, our knowledge of the satellite remains limited. Our ground-based photometry does not reveal any significant secondary periodic signal that can be attributed to the satellite, and the available \textit{HST} observations are too sparse to constrain its rotational period. However, the satellite exhibits a photometric variability of about 0.1–0.3~mag in the \textit{HST} data, suggesting possible low level albedo spots or a small to moderate elongation. If the satellite is elongated and does have a lightcurve amplitude of about 0.3~mag, the equilibrium figure models of \citet{Chandrasekhar1987} can be used to estimate a lower limit on its elongation. Under the assumption of an equator-on viewing geometry, a 0.3~mag amplitude corresponds to axis ratios of $b/a=0.76$ and $c/a=0.50$. Adopting a diameter of 141~km for the satellite, this yields approximate dimensions of $a=106$~km, $b=81$~km, and $c=40$~km \citep{Proudfoot2026}. To limit the range of plausible scenarios in our mutual event predictions (Section~\ref{sec:model}), we consider two cases: (1) a spherical satellite with a diameter of 141~km, and (2) a triaxial satellite with dimensions of 106$\times$81$\times$40~km. In both cases, we assume a density of 1.4~g~cm$^{-3}$ and an albedo of 0.17 \citep{Vilenius2012}. Since the satellite’s rotational period is unknown, we adopt a nominal (by default value, as the satellite's period is unconstrained) value of 3~days for illustrative purposes. These assumptions represent our current best estimate for the satellite, but additional observations will be needed to better constrain its shape and rotational properties.

Predictions based on the non-Keplerian orbit solution are presented in \citet{Proudfoot2026}. We emphasize that additional astrometric observations obtained closer to the mutual event season will be essential to refine the orbit, test the evidence for non-Keplerian motion, and improve the predicted timing of future mutual events. Once the first mutual event is observed, the system model can be further refined, reducing the timing uncertainties of subsequent events to only a few minutes as well as improving the determination of the date of the mutual events season end (see Section~\ref{sec:endseason}). With the current Keplerian orbit, the timing uncertainty of the individual mutual event is $\sim$5 (for inferior events) to $\sim$7~hours (for superior events).  
 
%%%%%%%%%%%%%%%%%%%%%%%
%%%%%%%%%%%%%
 \section{Mutual events season prediction} 
\label{sec:season}

 To predict the mutual events season, we use the system's model described in Section~\ref{sec:model} and the Keplerian mutual orbit derived in Section~\ref{sec:orbit}. As mentioned, there is already a published prediction using the non-Keplerian orbit, and we do not expect a perfect match between our Keplerian prediction and the non-Keplerian prediction by \citet{Proudfoot2026}. 
  
In Table~\ref{tab:events}, we summarize the predicted mutual events for the upcoming season. An inferior event occurs when the primary passes in front of the satellite, while a superior event occurs when the satellite passes in front of the primary. We emphasize that some predicted events, particularly near the beginning and end of the season, are highly sensitive to the assumed component sizes as well as to the rotational phase at the time of the event. In some cases, our nominal model predicts that the two bodies pass very close to one another without producing an event; these cases are classified as close approaches. However, if the components are larger than assumed, or if their rotational phases or orientations differ from our nominal model, these close approaches could instead result in detectable mutual events.

Following, we present some individual mutual events as examples. 

\subsection{Start of the mutual events season: 2031-2033}

According to our modeling, the mutual event season of 2001~XR$_{254}$ is expected to begin in September 2031, consistent with the prediction of \citet{Proudfoot2026}. However, the exact onset of the season is highly sensitive to the assumed sizes, shapes, rotational phases, and spin orientations of the components. As a result, the first predicted events may not occur if the bodies are smaller than assumed or have different orientations.

\begin{figure}
  \includegraphics[width=9cm,angle=0]{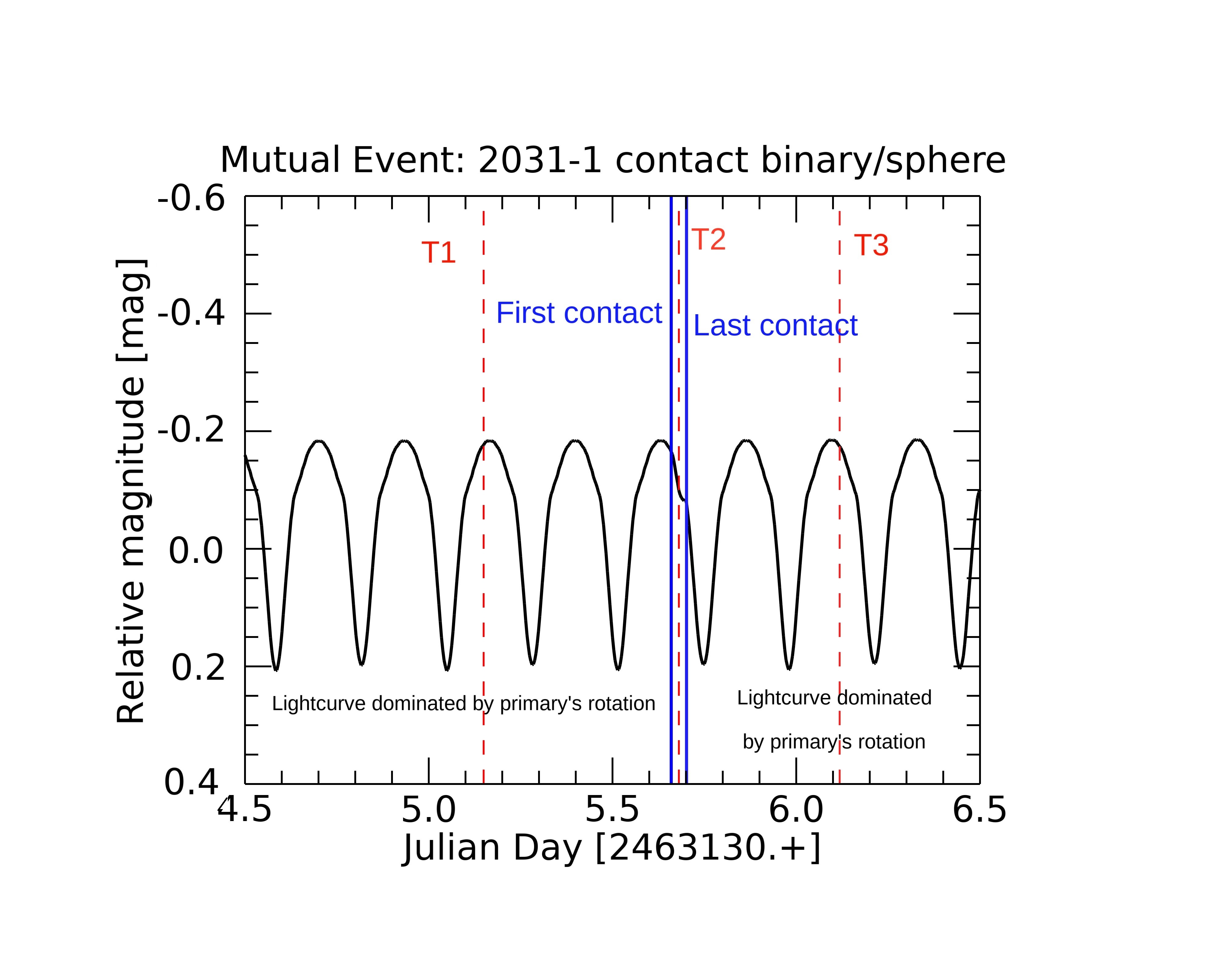} 
\includegraphics[width=9cm,angle=0]{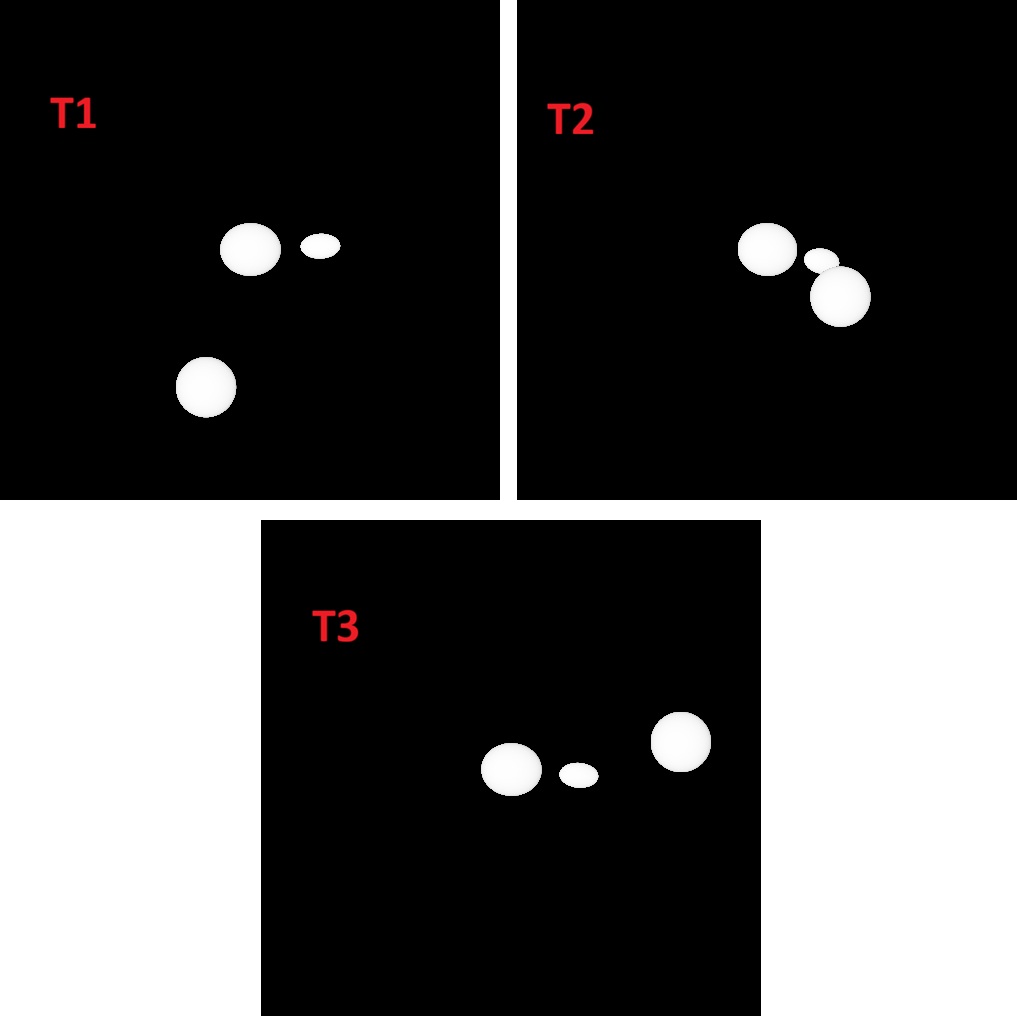} 
\caption{ We model the first predicted mutual event of the season under the assumption that the satellite is spherical. Because the satellite is represented as a uniform sphere with no surface brightness variations, its lightcurve is flat and its rotational period is irrelevant. In this configuration, the satellite passes in front of the smaller component of the primary, producing a grazing superior event. Owing to its grazing nature, the event is highly sensitive to the assumed satellite size and may not happen if the satellite is smaller than predicted.}
\label{fig:ME1a}
\end{figure}

\begin{figure}
  \includegraphics[width=9cm,angle=0]{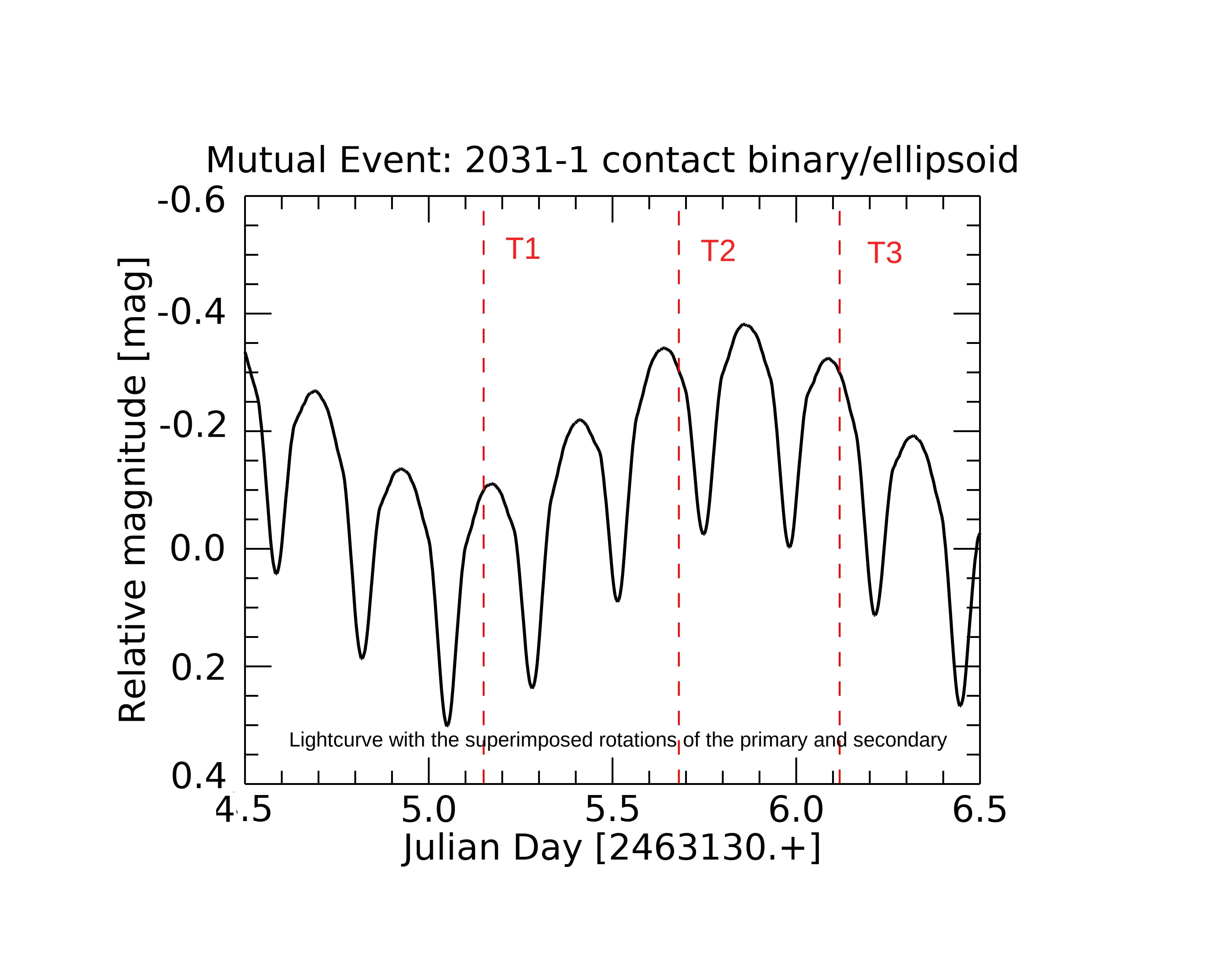} 
\includegraphics[width=9cm,angle=0]{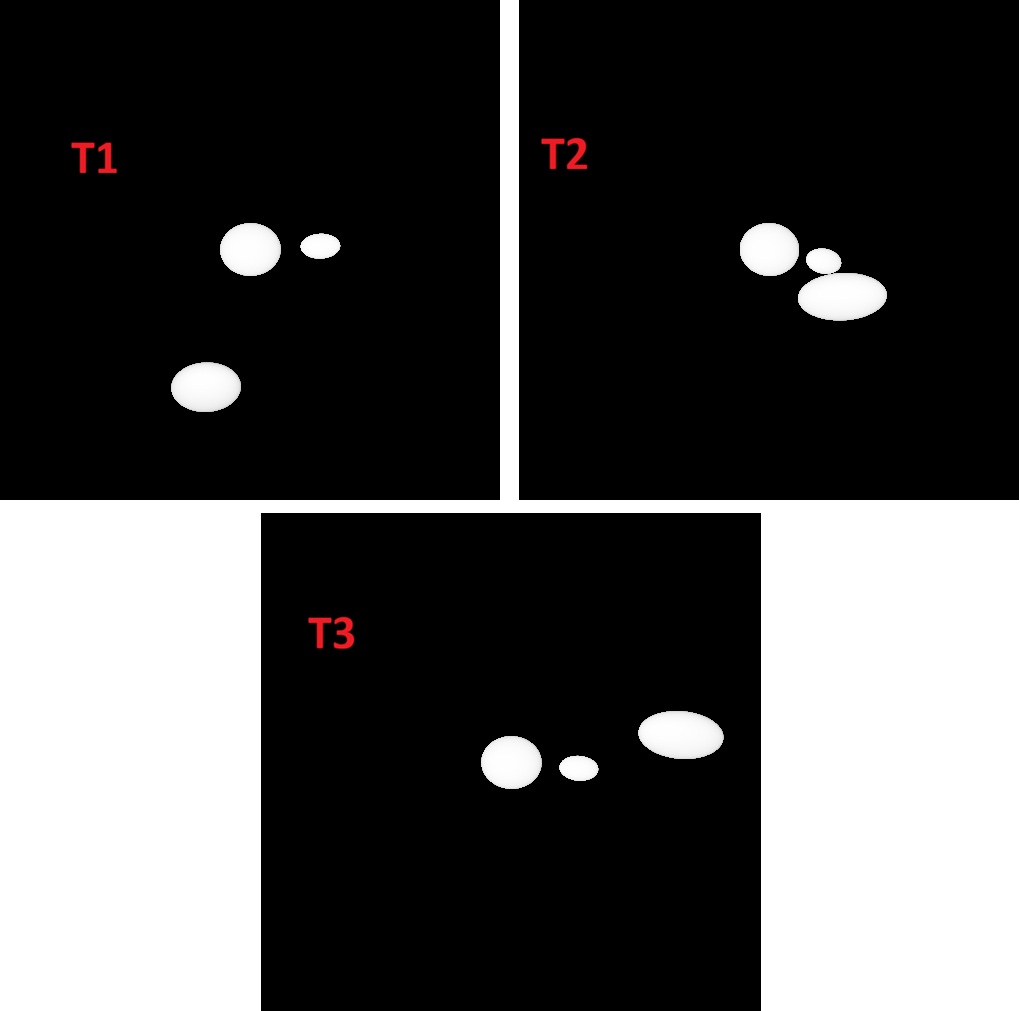} 
\caption{We also model the first predicted conjunction of the season assuming that the satellite is a triaxial ellipsoid. Because the satellite's rotational period is unknown, we adopt a nominal value of 3 days. Under these assumptions, no mutual event occurs and the components undergo only a close approach. However, this prediction is highly sensitive to the satellite's size, shape, and rotational phase (and thus period). Different physical properties or rotational states could shift the geometry sufficiently for a mutual event to occur.}
\label{fig:ME1b}
\end{figure}

Figures~\ref{fig:ME1a} and \ref{fig:ME1b} illustrate the first predicted event of the season under two different assumptions for the satellite shape: a sphere and a triaxial ellipsoid, respectively. In the spherical case (Figure~\ref{fig:ME1a}), the conjunction produces a grazing superior event in which the satellite partially occults the primary. A modest reduction in the satellite size, however, would be sufficient to eliminate the event. In contrast, when the satellite is modeled as an elongated ellipsoid (Figure~\ref{fig:ME1b}), the components undergo only a close approach and no mutual event occurs. These examples demonstrate that, although the mutual event season will likely begin in 2031, the exact timing of the first detectable event remains highly dependent on the object's sizes, rotational phases, and spin pole orientations. 

During the first years of the season, several conjunctions are predicted to result only in close approaches in our nominal model. Nevertheless, detectable mutual events could occur during some of these encounters if the component sizes, rotational phases, or spin pole orientations differ from our assumptions. In particular, events may occur near January~29,~2032, September~14,~2032, January~18,~2033, and February~9,~2033.

\subsection{Middle of the mutual events season: 2034-2037}

\begin{figure}
  \includegraphics[width=9cm,angle=0]{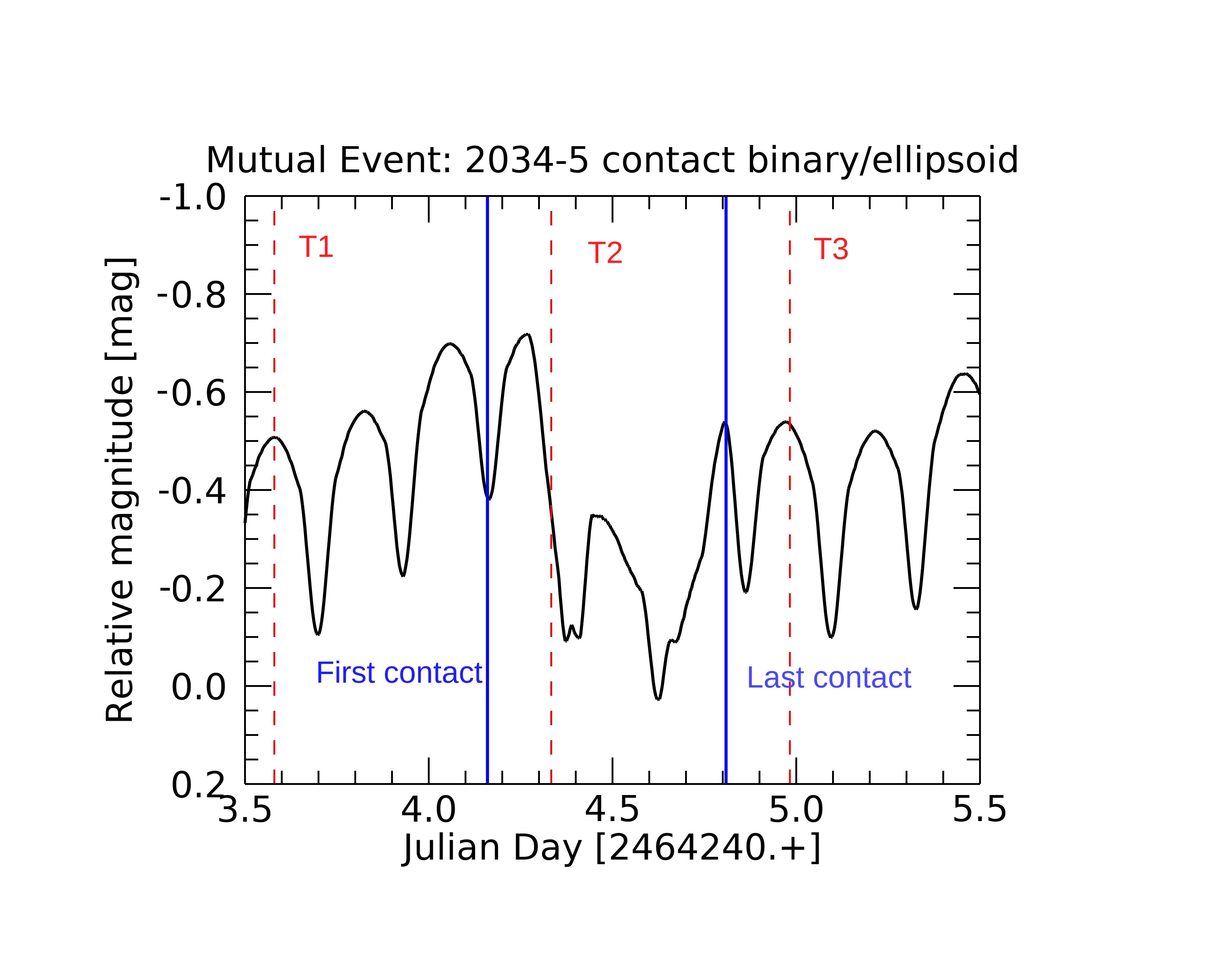} 
\includegraphics[width=9cm,angle=0]{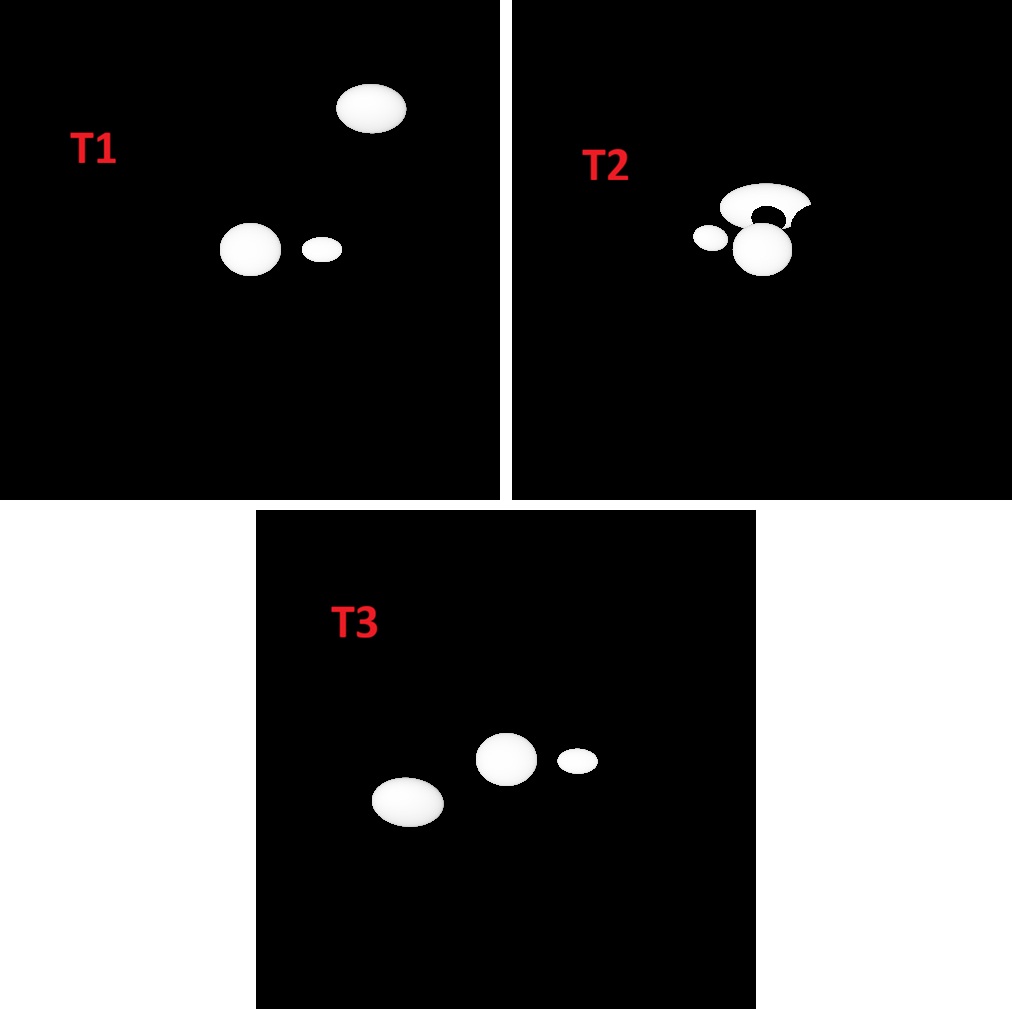}
\caption{We model the 2034-5 event assuming the satellite is a triaxial ellipsoid. In this case, the event is highly complex as there are multiple shadow casting into the satellite and objects nearly occulting others. }
\label{fig:ME2}
\end{figure}

 Up to 6 events per year are expected for this system at the midpoint of the mutual events season. Near mid-season, mutual events become deeper because the components are nearly, or even perfectly, aligned along the line of sight. However, these events can also be highly complex, involving multiple eclipses and occultations occurring in succession. As an example, Figure~\ref{fig:ME2} shows the simulated event 2034-5 assuming an ellipsoidal satellite. The event begins with the larger component of the primary eclipsing the satellite. As the satellite moves behind the primary and becomes partially occulted, the smaller component of the primary also starts to cast its shadow on the satellite. Finally, the smaller component partially occults the satellite. This event, therefore, involves both eclipses and occultations produced by both components of the primary, resulting in a particularly complex lightcurve (Figure~\ref{fig:ME2}).  

\begin{figure}
  \includegraphics[width=9cm,angle=0]{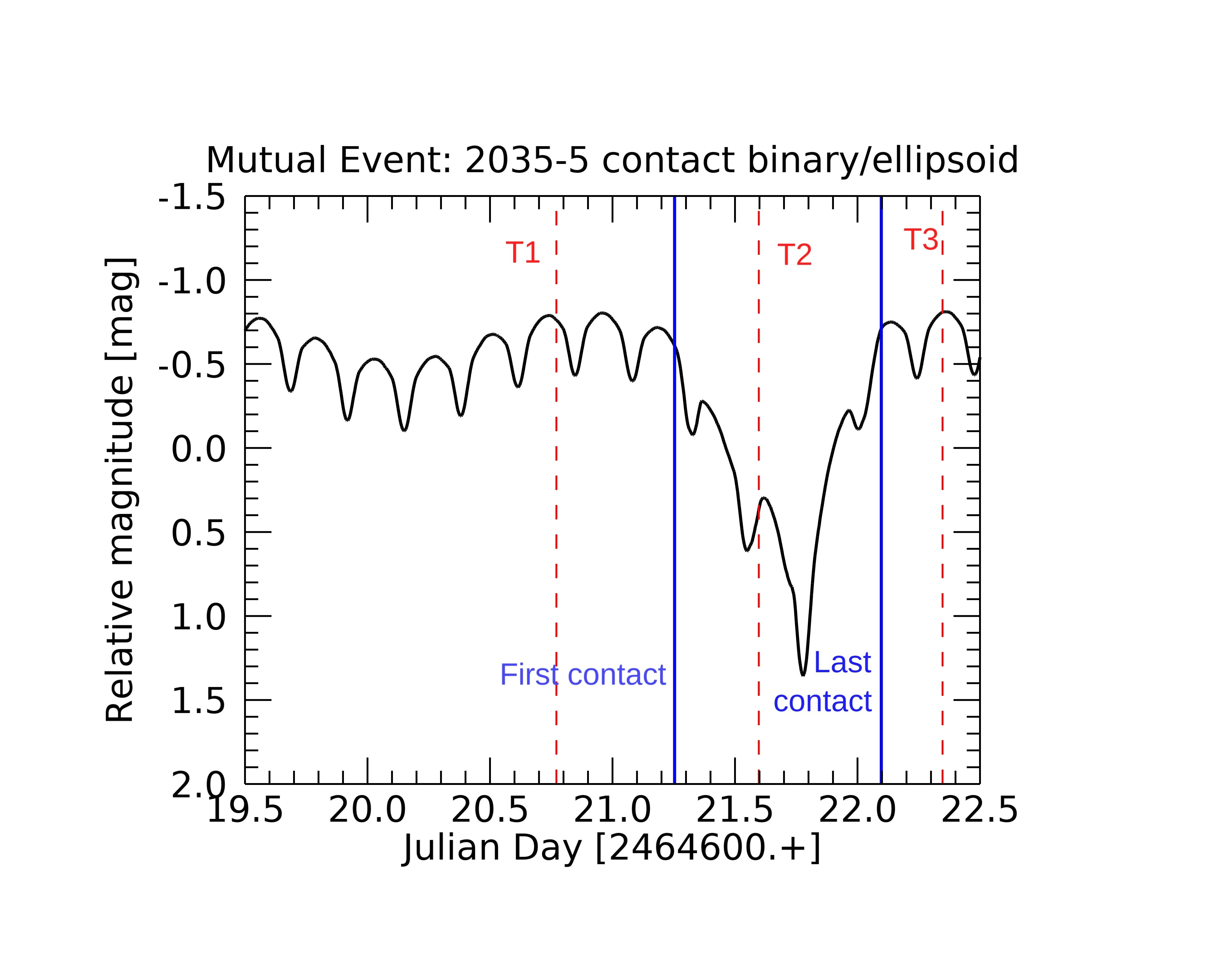} 
\includegraphics[width=9cm,angle=0]{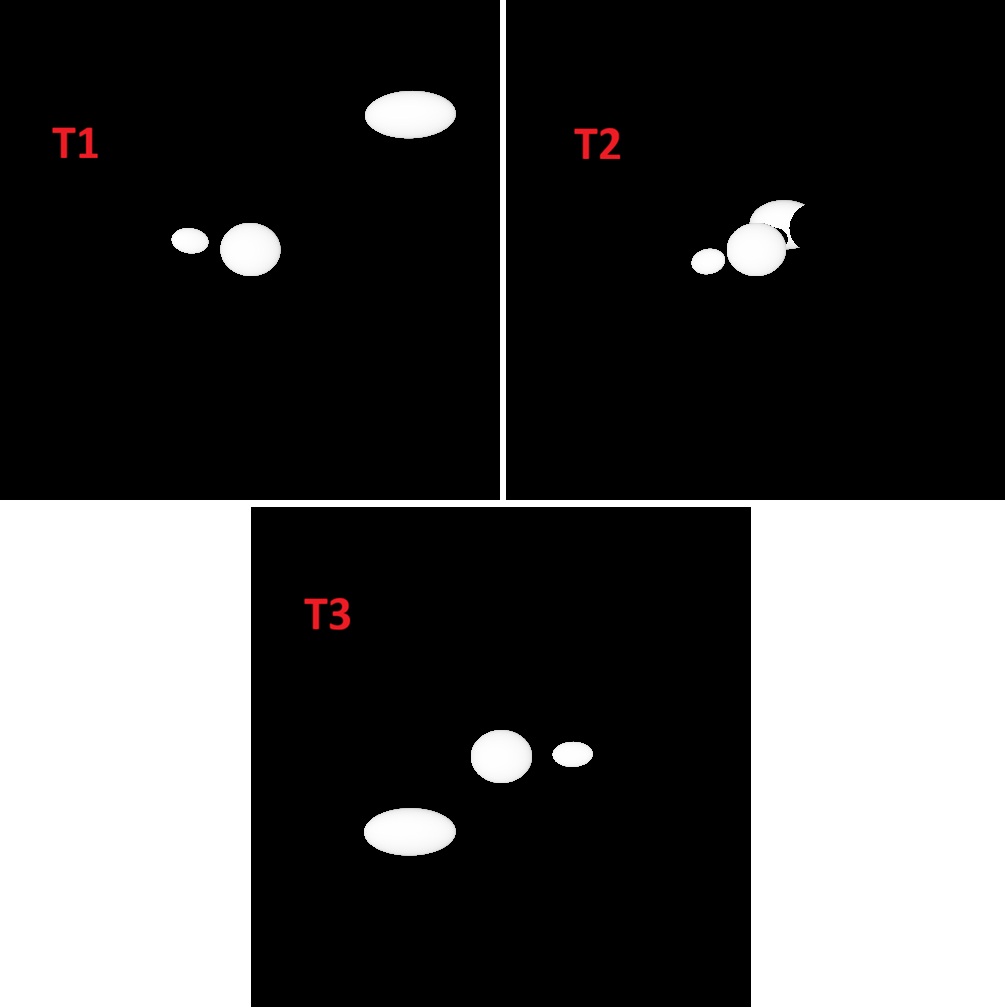}
\caption{As we reach the point of mid-season, the events are getting deeper. In this case, for the 2035-5 event, the primary repeatedly casts its shadow on the satellite, combined with a near-total occultation of the satellite by the primary.}
\label{fig:ME3}
\end{figure}

Another example is the 2035-5 event shown in Figure~\ref{fig:ME3}. This is one of the deepest events predicted during the mutual events season. The event consists of multiple eclipses, during which the primary repeatedly casts its shadow on the satellite, combined with a nearly total occultation of the satellite by the primary. The resulting lightcurve is both deep and complex.

\subsection{End of the mutual events season: 2038-2040}
\label{sec:endseason}

As at the beginning of the mutual events season, the timing of the final events is highly sensitive to the assumed sizes, shapes, rotational phases, and spin orientations of the components. In our nominal model, the last mutual event occurs in April 2040, but only if the satellite is represented as a triaxial ellipsoid. If the satellite is assumed to be spherical, the same conjunction results only in a close approach, and no mutual event is produced. We also identify a close approach on December~7,~2039, which does not generate a mutual event under either the spherical or ellipsoidal satellite models. These results illustrate the uncertainty surrounding the end of the mutual events season and highlight the importance of future observations for refining the system geometry and event predictions.

  \subsection{Keplerian versus non-Keplerian prediction}
  
Both the Keplerian model presented here and the non-Keplerian solution of \citet{Proudfoot2026} predict that the mutual-event season will begin in $\sim$2031 and extend until approximately 2040. Although the two models are not expected to agree perfectly, they produce broadly consistent predictions, including a similar number of events occurring at comparable times. Differences between the two solutions are typically limited to several hours in the predicted timing of individual events, which is well within expectations given the current uncertainties in the system architecture.

Aside from confirming the non-Keplerian nature of the system, a second rotational lightcurve acquired at a different epoch would help constrain the spin axis orientation of the primary, and improved characterization of the satellite's rotational and physical properties would significantly enhance future mutual events predictions.
    
 \section{Triple systems in the Kuiper belt}
 
 The high abundance of binary systems among dynamically Cold Classical KBOs is widely regarded as evidence that this population formed in a dynamically gentle environment and has undergone relatively little subsequent evolution \citep{Noll2020, Nesvorny2010}. However, the existence and diversity of triple systems (or hierarchical systems) remain largely unstudied. So far, one resolved triple system has been identified with the \textit{HST}: Lempo-Hiisi-Paha in the 3:2 mean motion resonance with Neptune \citep{Benecchi2010}. \citet{Rabinowitz2020} suggested that, in the 7:4 mean motion resonance, Manw{\"e} is a contact/close binary with a distant satellite named Thorondor. \citet{Nelsen2025} found evidence of non-Keplerian motion for Altjira, suggesting a potential triple system. \citet{Thirouin2025} also proposed that Logos-Zoe is a triple system or even a quadruple, and with this work, we suggest that 2001~XR$_{254}$ is also a triple system. Therefore, we are starting to probe the population of unresolved triple systems in the Kuiper belt. We emphasize that several of the mentioned systems (Altjira, Logos, and 2001~XR$_{254}$) are about or have already started their mutual events seasons, and thus we have a once-in-a-lifetime opportunity to model, predict, and observe these events to retrieve the multiplicity as well as physical and rotational properties of these systems.

%%%%%%%%%%%%%%%%%%%%%%%
%%%%%%%%%%%%%

 \section{Conclusion}
 
Using space- and ground-based observations of the resolved binary system 2001~XR$_{254}$, we find the following:

\begin{itemize}
\item We obtained the first ground-based rotational lightcurve of 2001~XR$_{254}$. The primary rotates with a period of 11.17$\pm$0.04~h and exhibits an asymmetric double-peaked lightcurve with an amplitude of 0.42$\pm$0.04~mag, one minimum being approximately 0.1~mag deeper than the other.

\item The primary's rotational period differs from the mutual orbital period, demonstrating that the system is asynchronous.

\item Lightcurve modeling suggests that the primary is a close/contact binary with a mass ratio of 0.2. For an assumed density of 1.4~g~cm$^{-3}$ and albedo of 0.17, the best-fit solution gives component~A dimensions of 142$\times$134$\times$118~km and component~B dimensions of 94$\times$72$\times$66~km with a separation of 169~km between the two components. We emphasize that this represents our current best-fit model, which can be refined with additional observations.

\item Our knowledge of the resolved satellite remains limited, as its rotational period is unknown and its shape is poorly constrained. Based on the available \textit{HST} data, we infer that it is likely either nearly spheroidal or modestly elongated.

\item We refined the Keplerian mutual orbit using \textit{HST} and \textit{Keck~II} astrometry acquired between 2006 and 2024 and used this solution to predict the upcoming mutual-event season. Our results indicate that the season should begin near the end of 2031 and continue until approximately 2040, in agreement with the non-Keplerian predictions of \citet{Proudfoot2026}.

\item We modeled the entire mutual events season and predicted at least 43 events over the next decade. The precise timing of the first and last events remains sensitive to the sizes, shapes, rotational states, and spin axis orientations of the components. Near mid-season, several events per year are expected as the system approaches near-perfect alignment, providing optimal opportunities for detailed characterization.

\item Our results suggest that 2001~XR$_{254}$ consists of a close/contact binary primary orbited by a distant satellite, making it a likely triple system. In this respect, it resembles the Cold Classical triple system Logos-Zoe, which is also approaching a mutual events season \citep{Thirouin2025}.

\item Recent work has revealed that multiple systems may be more common among Cold Classical KBOs than previously recognized \citep{Thirouin2025DPS}. If confirmed, 2001~XR$_{254}$ would join the growing population of known triple systems in the Kuiper Belt. Future observations, particularly during the upcoming mutual events season, will provide critical constraints on the architecture, formation, and evolution of this remarkable system. We will explore in more detail in an upcoming publication the multiple systems in the Kuiper belt. 

\end{itemize}
 
%%%%%%%%%%%%%%%%%%%%%%%
%%%%%%%%%%%%%

%% Please use the acknowledgment and contribution environments. This will 
%% be anonomyized when the "anonymous" style option is used. 

\begin{acknowledgments}

Authors thank the reviewer for useful comments and careful reading of the paper. 
 This research is based on data obtained at the Lowell Discovery Telescope (LDT). Lowell Observatory is a private, non-profit institution dedicated to astrophysical research and public appreciation of astronomy, and operates the LDT in partnership with Boston University, the University of Maryland, the University of Toledo, Northern Arizona University, and Yale University. Partial support of the LDT was provided by Discovery Communications. LMI was built by Lowell Observatory using funds from the National Science Foundation (AST-1005313). We are grateful to the  LDT staff.  \\ 
 Some of the data presented herein were obtained at Keck Observatory, 
which is a private 501(c)3 non-profit organization operated as a
scientific partnership among the California Institute of Technology,
the University of California, and the National Aeronautics and Space
Administration. The Observatory was made possible by the generous
financial support of the W. M. Keck Foundation.  The data are publicly
available through Keck Observatory Archive (KOA), which is operated by the
W. M. Keck Observatory and the NASA Exoplanet Science Institute (NExScI),
under contract with the National Aeronautics and Space Administration.
The authors wish to recognize and acknowledge the very significant
cultural role and reverence that the summit of Maunakea has always had
within the Native Hawaiian community. We are most fortunate to have the
opportunity to conduct observations from this mountain.
This research is based on observations made with the NASA/ESA Hubble 
Space Telescope obtained from the Space Telescope Science Institute,
which is operated by the Association of Universities for Research in
Astronomy, Inc., under NASA contract NAS 5–26555. These observations
are associated with programs \#10800, \#11178, and \#17707, with data hosted at 
the Multimission Archive at the Space Telescope Science Institute (MAST).\\
Authors made use of the JPL Horizons Systems, which was developed at the Jet Propulsion Laboratory (Solar System Dynamics Group), California Institute of Technology, under contract with the National Aeronautics and Space Administration.\\ 
AT, WMG, SSS, and KSN acknowledge support from NASA-Solar System Observations (SSO) with grant 80NSSC22K0659 awarded to “Mutual events in the trans-Neptunian belt”.
JMGL acknowledges financial support from the Spanish Ministry of Universities through the university training programme FPU2022/00492 and from the Severo Ochoa grant CEX2021-001131-S funded by MCIN/AEI/10.13039/501100011033.
\end{acknowledgments}

\begin{contribution}
%%This section gives authors the space to recognize author contributions. The text inside this environment is NOT counted towards the total word quanta. At a minimum, manuscripts are expected to include this text:

All authors contributed to this paper.

%% But authors are expected to provide more specific details, e.g. 
%%
%%SC was responsible for writing and submitting the manuscript.
%%WWM came up with the initial research concept and edited the manuscript.
%%OTS obtained the funding and edited the manuscript.
%%EBF provided the formal analysis and validation. He also edited the manuscript.
%%GEH Supervised the undergraduates, wrote the software and administers the project github and Zenodo repositories.
%%
%% Authors can use the Contributor Role Taxonomy (CRediT) at
%% for ideas on how write a good statement tailored to their needs.

\end{contribution}

\facilities{LDT, HST, Keck:II}

%%%%%\software{astropy \citep{2013A&A...558A..33A,2018AJ....156..123A,2022ApJ...935..167A},  Cloudy \citep{2013RMxAA..49..137F}, Source Extractor \citep{1996A&AS..117..393B}}


  \clearpage
 \begin{longtable}{|l|l|l|l|l|}
\caption{\label{tab:events} Predicted mutual events for 2001~XR$_{254}$.}\\
\hline
Event & Shapes primary/satellite & First contact [JD] &  Last contact [JD] & Note \\
\hline
\endfirsthead

\hline
Event & Shapes primary/satellite & First contact [JD] &  Last contact [JD] & Note \\
\hline
\endhead

\hline
\endfoot

\hline
\endlastfoot
\multicolumn{5}{|c|}{Year: 2031}                                                                                                                   \\ \hline
\multicolumn{1}{|l|}{2031-1} & \multicolumn{1}{l|}{contact binary/sphere} & \multicolumn{1}{l|}{2463135.65972}      & \multicolumn{1}{l|}{2463135.70139}    & superior \\ \hline
\multicolumn{1}{|l|}{2031-1} & \multicolumn{1}{l|}{contact binary/ellipsoid} & \multicolumn{1}{l|}{}      & \multicolumn{1}{l|}{}    & close approach \\ \hline
\multicolumn{5}{|c|}{Year: 2032}                                                                                                                   \\ \hline
\multicolumn{1}{|l|}{2032-1}       & \multicolumn{1}{l|}{contact binary/sphere}     & \multicolumn{1}{l|}{2463386.94097}      & \multicolumn{1}{l|}{2463386.97569}    &   superior \\ \hline
\multicolumn{1}{|l|}{2032-1}       & \multicolumn{1}{l|}{contact binary/ellipsoid}                      & \multicolumn{1}{l|}{ }      & \multicolumn{1}{l|}{}    & close approach  \\ \hline 
\multicolumn{1}{|l|}{2032-2}       & \multicolumn{1}{l|}{contact binary/sphere}                      & \multicolumn{1}{l|}{2463512.59722}      & \multicolumn{1}{l|}{2463513.04514}    &   superior            \\ \hline
\multicolumn{1}{|l|}{2032-2}       & \multicolumn{1}{l|}{contact binary/ellipsoid}                      & \multicolumn{1}{l|}{2463512.73958}      & \multicolumn{1}{l|}{2463513.04861 }    &    \\ \hline  
\multicolumn{5}{|c|}{Year: 2033}                                                                                                                   \\ \hline
\multicolumn{1}{|l|}{2033-1}       & \multicolumn{1}{l|}{contact binary/sphere}    & \multicolumn{1}{l|}{2463763.87153}      & \multicolumn{1}{l|}{2463764.31944}    &    superior            \\ \hline
\multicolumn{1}{|l|}{2033-1}       & \multicolumn{1}{l|}{contact binary/ellipsoid}                 & \multicolumn{1}{l|}{2463764.22222}      & \multicolumn{1}{l|}{2463764.32986}    &                \\ \hline
\multicolumn{1}{|l|}{2033-2}       & \multicolumn{1}{l|}{contact binary/sphere}    & \multicolumn{1}{l|}{2463867.29861}      & \multicolumn{1}{l|}{2463867.44097}    &    inferior            \\ \hline
\multicolumn{1}{|l|}{2033-2}       & \multicolumn{1}{l|}{contact binary/ellipsoid}      & \multicolumn{1}{l|}{2463867.31250 }      & \multicolumn{1}{l|}{2463867.44792}    &                \\ \hline
\multicolumn{1}{|l|}{2033-3}       & \multicolumn{1}{l|}{contact binary/sphere}    & \multicolumn{1}{l|}{2463889.67708}      & \multicolumn{1}{l|}{2463890.02431}    &    superior            \\ \hline
\multicolumn{1}{|l|}{2033-3}       & \multicolumn{1}{l|}{contact binary/ellipsoid}    & \multicolumn{1}{l|}{2463889.73611}      & \multicolumn{1}{l|}{2463890.07639}    &                 \\ \hline
\multicolumn{5}{|c|}{Year: 2034}                                                                                                                   \\ \hline
\multicolumn{1}{|l|}{2034-1}       & \multicolumn{1}{l|}{contact binary/sphere}                        & \multicolumn{1}{l|}{2463992.93056}      & \multicolumn{1}{l|}{2463993.05556}    &        inferior        \\ \hline
\multicolumn{1}{|l|}{2034-1}       & \multicolumn{1}{l|}{contact binary/ellipsoid}    & \multicolumn{1}{l|}{2463992.93750}      & \multicolumn{1}{l|}{2463993.05903}    &                 \\ \hline
\multicolumn{1}{|l|}{2034-2}       & \multicolumn{1}{l|}{contact binary/sphere}                        & \multicolumn{1}{l|}{2464015.28819}      & \multicolumn{1}{l|}{2464015.62847}    &        superior        \\ \hline
\multicolumn{1}{|l|}{2034-2}       & \multicolumn{1}{l|}{contact binary/ellipsoid}    & \multicolumn{1}{l|}{2464015.31250}      & \multicolumn{1}{l|}{2464015.63194}    &                 \\ \hline
\multicolumn{1}{|l|}{2034-3}       & \multicolumn{1}{l|}{contact binary/sphere}                        & \multicolumn{1}{l|}{2464118.57986}      & \multicolumn{1}{l|}{2464118.71528}    &        inferior        \\ \hline
\multicolumn{1}{|l|}{2034-3}       & \multicolumn{1}{l|}{contact binary/ellipsoid}    & \multicolumn{1}{l|}{2464118.59375}      & \multicolumn{1}{l|}{2464118.72222}    &                 \\ \hline
\multicolumn{1}{|l|}{2034-4}       & \multicolumn{1}{l|}{contact binary/sphere}                        & \multicolumn{1}{l|}{2464140.93403}      & \multicolumn{1}{l|}{2464141.29861}    &        superior        \\ \hline
\multicolumn{1}{|l|}{2034-4}       & \multicolumn{1}{l|}{contact binary/ellipsoid}    & \multicolumn{1}{l|}{2464140.96181}      & \multicolumn{1}{l|}{2464141.30208}    &                 \\ \hline
\multicolumn{1}{|l|}{2034-5}       & \multicolumn{1}{l|}{contact binary/sphere}                        & \multicolumn{1}{l|}{2464244.09375}      & \multicolumn{1}{l|}{2464244.81250}    &        inferior        \\ \hline
\multicolumn{1}{|l|}{2034-5}       & \multicolumn{1}{l|}{contact binary/ellipsoid}    & \multicolumn{1}{l|}{2464244.15972}      & \multicolumn{1}{l|}{2464244.81597}    &                 \\ \hline
\multicolumn{1}{|l|}{2034-6}       & \multicolumn{1}{l|}{contact binary/sphere}                        & \multicolumn{1}{l|}{2464266.65278}      & \multicolumn{1}{l|}{2464267.31944}    &        superior        \\ \hline
\multicolumn{1}{|l|}{2034-6}       & \multicolumn{1}{l|}{contact binary/ellipsoid}    & \multicolumn{1}{l|}{2464244.15972}      & \multicolumn{1}{l|}{2464244.81597}    &                 \\ \hline
\multicolumn{5}{|c|}{Year: 2035}                                                                                                                   \\ \hline
\multicolumn{1}{|l|}{2035-1}       & \multicolumn{1}{l|}{contact binary/sphere}                        & \multicolumn{1}{l|}{2464369.77083}      & \multicolumn{1}{l|}{2464370.40278}    &        inferior        \\ \hline
\multicolumn{1}{|l|}{2035-1}       & \multicolumn{1}{l|}{contact binary/ellipsoid}    & \multicolumn{1}{l|}{2464369.82292}      & \multicolumn{1}{l|}{2464370.39931}    &                 \\ \hline
\multicolumn{1}{|l|}{2035-2}       & \multicolumn{1}{l|}{contact binary/sphere}                        & \multicolumn{1}{l|}{2464392.28472}      & \multicolumn{1}{l|}{2464392.70486}    &        superior        \\ \hline
\multicolumn{1}{|l|}{2035-2}       & \multicolumn{1}{l|}{contact binary/ellipsoid}    & \multicolumn{1}{l|}{2464392.31597}      & \multicolumn{1}{l|}{2464392.75694}    &                 \\ \hline
\multicolumn{1}{|l|}{2035-3}       & \multicolumn{1}{l|}{contact binary/sphere}                        & \multicolumn{1}{l|}{2464495.36458 }      & \multicolumn{1}{l|}{2464496.07986 }    &        inferior        \\ \hline
\multicolumn{1}{|l|}{2035-3}       & \multicolumn{1}{l|}{contact binary/ellipsoid}    & \multicolumn{1}{l|}{2464495.40625 }      & \multicolumn{1}{l|}{2464496.07986 }    &                 \\ \hline
\multicolumn{1}{|l|}{2035-4}       & \multicolumn{1}{l|}{contact binary/sphere}                        & \multicolumn{1}{l|}{ 2464517.95486 }      & \multicolumn{1}{l|}{ 2464518.58681 }    &        superior        \\ \hline
\multicolumn{1}{|l|}{2035-4}       & \multicolumn{1}{l|}{contact binary/ellipsoid}    & \multicolumn{1}{l|}{2464517.98264 }      & \multicolumn{1}{l|}{2464518.57639 }    &                 \\ \hline
\multicolumn{1}{|l|}{2035-5}       & \multicolumn{1}{l|}{contact binary/sphere}                        & \multicolumn{1}{l|}{2464621.22569 }      & \multicolumn{1}{l|}{2464622.11458 }    &        inferior        \\ \hline
\multicolumn{1}{|l|}{2035-5}       & \multicolumn{1}{l|}{contact binary/ellipsoid}    & \multicolumn{1}{l|}{2464621.25347}      & \multicolumn{1}{l|}{2464622.09722 }    &                 \\ \hline
\multicolumn{1}{|l|}{2035-6}       & \multicolumn{1}{l|}{contact binary/sphere}                        & \multicolumn{1}{l|}{2464643.73264 }      & \multicolumn{1}{l|}{2464644.35069 }    &        superior        \\ \hline
\multicolumn{1}{|l|}{2035-6}       & \multicolumn{1}{l|}{contact binary/ellipsoid}    & \multicolumn{1}{l|}{2464643.75000 }      & \multicolumn{1}{l|}{2464644.35417 }    &                 \\ \hline
\multicolumn{5}{|c|}{Year: 2036}                                                                                                                   \\ \hline
\multicolumn{1}{|l|}{2036-1}       & \multicolumn{1}{l|}{contact binary/sphere}                        & \multicolumn{1}{l|}{2464746.89931}      & \multicolumn{1}{l|}{2464747.50694}    &        inferior        \\ \hline
\multicolumn{1}{|l|}{2036-1}       & \multicolumn{1}{l|}{contact binary/ellipsoid}    & \multicolumn{1}{l|}{2464746.92014}      & \multicolumn{1}{l|}{2464747.51736}    &                 \\ \hline
\multicolumn{1}{|l|}{2036-2}       & \multicolumn{1}{l|}{contact binary/sphere}                        & \multicolumn{1}{l|}{2464769.33681}      & \multicolumn{1}{l|}{2464769.89931}    &        superior        \\ \hline
\multicolumn{1}{|l|}{2036-2}       & \multicolumn{1}{l|}{contact binary/ellipsoid}    & \multicolumn{1}{l|}{2464769.35417}      & \multicolumn{1}{l|}{2464769.88194}    &                 \\ \hline
\multicolumn{1}{|l|}{2036-3}       & \multicolumn{1}{l|}{contact binary/sphere}               & \multicolumn{1}{l|}{2464872.48611}      & \multicolumn{1}{l|}{2464873.38542}    &       inferior         \\ \hline 
\multicolumn{1}{|l|}{2036-3}       & \multicolumn{1}{l|}{contact binary/ellipsoid}            & \multicolumn{1}{l|}{2464872.50694}      & \multicolumn{1}{l|}{2464873.36806}    &                \\ \hline
\multicolumn{1}{|l|}{2036-4}       & \multicolumn{1}{l|}{contact binary/sphere}               & \multicolumn{1}{l|}{2464895.06597}      & \multicolumn{1}{l|}{2464895.57986}    &       superior         \\ \hline 
\multicolumn{1}{|l|}{2036-4}       & \multicolumn{1}{l|}{contact binary/ellipsoid}            & \multicolumn{1}{l|}{2464895.06597}      & \multicolumn{1}{l|}{2464895.59028}    &                \\ \hline
\multicolumn{1}{|l|}{2036-5}       & \multicolumn{1}{l|}{contact binary/sphere}               & \multicolumn{1}{l|}{2464998.46181 }      & \multicolumn{1}{l|}{ 2464999.30208}    &       inferior         \\ \hline 
\multicolumn{1}{|l|}{2036-5}       & \multicolumn{1}{l|}{contact binary/ellipsoid}            & \multicolumn{1}{l|}{2464998.44792 }      & \multicolumn{1}{l|}{2464999.26736 }    &                \\ \hline
\multicolumn{1}{|l|}{2036-6}       & \multicolumn{1}{l|}{contact binary/sphere}               & \multicolumn{1}{l|}{2465020.80903 }      & \multicolumn{1}{l|}{ 2465021.46181}    &       superior         \\ \hline 
\multicolumn{1}{|l|}{2036-6}       & \multicolumn{1}{l|}{contact binary/ellipsoid}            & \multicolumn{1}{l|}{2465020.80556 }      & \multicolumn{1}{l|}{2465021.44444 }    &                \\ \hline
\multicolumn{5}{|c|}{Year: 2037}                                                                                                                   \\ \hline
\multicolumn{1}{|l|}{2037-1}       & \multicolumn{1}{l|}{contact binary/sphere}               & \multicolumn{1}{l|}{2465124.08333 }      & \multicolumn{1}{l|}{2465124.76389  }    &       inferior         \\ \hline 
\multicolumn{1}{|l|}{2037-1}       & \multicolumn{1}{l|}{contact binary/ellipsoid}            & \multicolumn{1}{l|}{2465124.08333 }      & \multicolumn{1}{l|}{2465124.76389  }     &                \\ \hline
\multicolumn{1}{|l|}{2037-2}       & \multicolumn{1}{l|}{contact binary/sphere}               & \multicolumn{1}{l|}{2465146.42014 }      & \multicolumn{1}{l|}{2465146.99306  }    &       superior         \\ \hline 
\multicolumn{1}{|l|}{2037-2}       & \multicolumn{1}{l|}{contact binary/ellipsoid}            & \multicolumn{1}{l|}{ 2465146.42708}      & \multicolumn{1}{l|}{2465146.99306 }    &                \\ \hline
\multicolumn{1}{|l|}{2037-3}       & \multicolumn{1}{l|}{contact binary/sphere}               & \multicolumn{1}{l|}{ 2465249.77083}      & \multicolumn{1}{l|}{ 2465250.51736 }    &       inferior         \\ \hline 
\multicolumn{1}{|l|}{2037-3}       & \multicolumn{1}{l|}{contact binary/ellipsoid}            & \multicolumn{1}{l|}{ 2465249.76736}      & \multicolumn{1}{l|}{2465250.50347 }    &                \\ \hline
\multicolumn{1}{|l|}{2037-4}       & \multicolumn{1}{l|}{contact binary/sphere}               & \multicolumn{1}{l|}{2465272.12153 }      & \multicolumn{1}{l|}{ 2465272.70486 }    &       superior         \\ \hline 
\multicolumn{1}{|l|}{2037-4}       & \multicolumn{1}{l|}{contact binary/ellipsoid}            & \multicolumn{1}{l|}{2465272.12500 }      & \multicolumn{1}{l|}{2465272.69444 }    &                \\ \hline
\multicolumn{1}{|l|}{2037-5}       & \multicolumn{1}{l|}{contact binary/sphere}               & \multicolumn{1}{l|}{2465375.64236 }      & \multicolumn{1}{l|}{2465376.43750  }    &       inferior         \\ \hline 
\multicolumn{1}{|l|}{2037-5}       & \multicolumn{1}{l|}{contact binary/ellipsoid}            & \multicolumn{1}{l|}{ 2465375.63194}      & \multicolumn{1}{l|}{2465376.36111 }    &                \\ \hline
\multicolumn{1}{|l|}{2037-6}       & \multicolumn{1}{l|}{contact binary/sphere}               & \multicolumn{1}{l|}{ 2465397.99306}      & \multicolumn{1}{l|}{2465398.53819  }    &       superior         \\ \hline 
\multicolumn{1}{|l|}{2037-6}       & \multicolumn{1}{l|}{contact binary/ellipsoid}            & \multicolumn{1}{l|}{ 2465397.95139}      & \multicolumn{1}{l|}{2465398.50347 }    &                \\ \hline 
\multicolumn{5}{|c|}{Year: 2038}                                                                                                                   \\ \hline
\multicolumn{1}{|l|}{2038-1}       & \multicolumn{1}{l|}{contact binary/sphere}               & \multicolumn{1}{l|}{2465501.26042 }      & \multicolumn{1}{l|}{ 2465501.97222 }    &       inferior         \\ \hline 
\multicolumn{1}{|l|}{2038-1}       & \multicolumn{1}{l|}{contact binary/ellipsoid}            & \multicolumn{1}{l|}{ 2465501.25000}      & \multicolumn{1}{l|}{ 2465501.93056}    &                \\ \hline
\multicolumn{1}{|l|}{2038-2}       & \multicolumn{1}{l|}{contact binary/sphere}               & \multicolumn{1}{l|}{2465523.40625 }      & \multicolumn{1}{l|}{ 2465524.09722 }    &       superior         \\ \hline 
\multicolumn{1}{|l|}{2038-2}       & \multicolumn{1}{l|}{contact binary/ellipsoid}            & \multicolumn{1}{l|}{ 2465523.50347}      & \multicolumn{1}{l|}{2465524.07639 }    &                \\ \hline
\multicolumn{1}{|l|}{2038-3}       & \multicolumn{1}{l|}{contact binary/sphere}               & \multicolumn{1}{l|}{2465626.94444 }      & \multicolumn{1}{l|}{2465627.67708  }    &       inferior         \\ \hline 
\multicolumn{1}{|l|}{2038-3}       & \multicolumn{1}{l|}{contact binary/ellipsoid}            & \multicolumn{1}{l|}{ 2465626.93403}      & \multicolumn{1}{l|}{2465627.62500 }    &                \\ \hline
\multicolumn{1}{|l|}{2038-4}       & \multicolumn{1}{l|}{contact binary/sphere}               & \multicolumn{1}{l|}{ 2465649.43750}      & \multicolumn{1}{l|}{ 2465649.79514 }    &       superior         \\ \hline 
\multicolumn{1}{|l|}{2038-4}       & \multicolumn{1}{l|}{contact binary/ellipsoid}            & \multicolumn{1}{l|}{2465649.39931 }      & \multicolumn{1}{l|}{2465649.75694 }    &                \\ \hline
\multicolumn{1}{|l|}{2038-5}       & \multicolumn{1}{l|}{contact binary/sphere}               & \multicolumn{1}{l|}{2465753.02778 }      & \multicolumn{1}{l|}{ 2465753.34722 }    &       inferior         \\ \hline 
\multicolumn{1}{|l|}{2038-5}       & \multicolumn{1}{l|}{contact binary/ellipsoid}            & \multicolumn{1}{l|}{ 2465753.02083}      & \multicolumn{1}{l|}{ 2465753.28819}    &                \\ \hline
\multicolumn{1}{|l|}{2038-6}       & \multicolumn{1}{l|}{contact binary/sphere}               & \multicolumn{1}{l|}{ 2465775.12153}      & \multicolumn{1}{l|}{ 2465775.47569 }    &       superior         \\ \hline 
\multicolumn{1}{|l|}{2038-6}       & \multicolumn{1}{l|}{contact binary/ellipsoid}            & \multicolumn{1}{l|}{ 2465775.11458}      & \multicolumn{1}{l|}{ 2465775.43403}    &                \\ \hline  
\multicolumn{5}{|c|}{Year: 2039}                                                                                                                   \\ \hline
\multicolumn{1}{|l|}{2039-1}       & \multicolumn{1}{l|}{contact binary/sphere}               & \multicolumn{1}{l|}{ 2465878.60069}      & \multicolumn{1}{l|}{ 2465878.95139 }    &       inferior         \\ \hline 
\multicolumn{1}{|l|}{2039-1}       & \multicolumn{1}{l|}{contact binary/ellipsoid}            & \multicolumn{1}{l|}{2465878.56250 }      & \multicolumn{1}{l|}{ 2465878.89931}    &                \\ \hline
\multicolumn{1}{|l|}{2039-2}       & \multicolumn{1}{l|}{contact binary/sphere}               & \multicolumn{1}{l|}{2465900.67361 }      & \multicolumn{1}{l|}{ 2465901.14583 }    &       superior         \\ \hline 
\multicolumn{1}{|l|}{2039-2}       & \multicolumn{1}{l|}{contact binary/ellipsoid}            & \multicolumn{1}{l|}{2465900.62500 }      & \multicolumn{1}{l|}{2465901.10069 }    &                \\ \hline
\multicolumn{1}{|l|}{2039-3}       & \multicolumn{1}{l|}{contact binary/sphere}               & \multicolumn{1}{l|}{2466004.33333 }      & \multicolumn{1}{l|}{ 2466004.47222 }    &       inferior         \\ \hline 
\multicolumn{1}{|l|}{2039-3}       & \multicolumn{1}{l|}{contact binary/ellipsoid}            & \multicolumn{1}{l|}{ 2466004.32292}      & \multicolumn{1}{l|}{2466004.48264 }    &                \\ \hline
\multicolumn{1}{|l|}{2039-4}       & \multicolumn{1}{l|}{contact binary/sphere}               & \multicolumn{1}{l|}{ 2466026.44444}      & \multicolumn{1}{l|}{ 2466026.57292 }    &       superior         \\ \hline 
\multicolumn{1}{|l|}{2039-4}       & \multicolumn{1}{l|}{contact binary/ellipsoid}            & \multicolumn{1}{l|}{ 2466026.46181}      & \multicolumn{1}{l|}{2466026.76736 }    &                \\ \hline
\multicolumn{1}{|l|}{2039-5}       & \multicolumn{1}{l|}{contact binary/sphere}               & \multicolumn{1}{l|}{ 2466152.10069}      & \multicolumn{1}{l|}{ 2466152.16667 }    &       superior         \\ \hline 
\multicolumn{1}{|l|}{2039-5}       & \multicolumn{1}{l|}{contact binary/ellipsoid}            & \multicolumn{1}{l|}{2466152.08333 }      & \multicolumn{1}{l|}{ 2466152.19097}    &                \\ \hline
\multicolumn{5}{|c|}{Year: 2040}                                                                                                                   \\ \hline
\multicolumn{1}{|l|}{2040-1}       & \multicolumn{1}{l|}{contact binary/sphere}               & \multicolumn{1}{l|}{ }      & \multicolumn{1}{l|}{  }    &   close approach   \\ \hline 
\multicolumn{1}{|l|}{2040-1}       & \multicolumn{1}{l|}{contact binary/ellipsoid}            & \multicolumn{1}{l|}{2466255.60069 }      & \multicolumn{1}{l|}{2466255.74653 }    &         inferior       \\ \hline
% \end{tabular}
\end{longtable}

\newpage

\bibliography{biblio}{}
\bibliographystyle{aasjournalv7}

\end{document}